\documentclass[superscriptaddress,aps,pra,twocolumn,floatfix]{revtex4-2}

\usepackage{graphicx}% Include figure files
\usepackage{dcolumn}% Align table columns on decimal point
\usepackage{bm}% bold math
\usepackage{multirow}%
\usepackage{amsmath,amssymb,amsfonts}%
\usepackage{amsthm}%
\usepackage{mathrsfs}%
\usepackage{bbm}%
\usepackage{stackengine}
 \usepackage[title]{appendix}%
 \usepackage{xcolor}%
 \usepackage{textcomp}%
 \usepackage{booktabs}%
 \usepackage{algorithm}%
 \usepackage{algorithmicx}%
 \usepackage{algpseudocode}%
 \usepackage{listings}%
\usepackage{hyperref}%
\usepackage{xurl}
\usepackage{color}
\definecolor{mygreen}{rgb}{0,0.5,0}
\definecolor{mygrey}{rgb}{0.5,0.5,0.5}
\definecolor{myred}{rgb}{0.75,0,0}
\definecolor{myblue}{rgb}{0,0,0.75}
\definecolor{mymagenta}{cmyk}{0,1,0,0.12}
\definecolor{mycyan}{cmyk}{1,0,0,0.12}
\definecolor{myorange}{rgb}{1.,0.5,0}
\definecolor{myviolet}{rgb}{0.6,0.15,0.6}
\definecolor{mybrown}{cmyk}{0,0.50,1,0.41}
\usepackage[normalem]{ulem}

\newcommand{\bF}{{\bf F}}
\newcommand{\bB}{{\bf B}}

\newcommand{\bN}{{\bf N}}

\newcommand{\bFmax}{{\mathbf{F}_{\rm max}}}

\newcommand{\supin}{^{({\rm in})}}
\newcommand{\supout}{^{({\rm out})}}

\newcommand{\rfsym}{\rf}
\newcommand{\subdc}{_\mathrm{dc}}
\newcommand{\subrf}{_\mathrm{rf}}
\newcommand{\bdc}{B\subdc}

\newcommand{\bvdc}{\mathbf{B}\subdc}

\newcommand{\brf}{B\subrf}
\newcommand{\bvrf}{\mathbf{B}\subrf}

\newcommand{\qsym}{\mathcal{Q}}
\newcommand{\isym}{\mathcal{I}}

\newcommand{\subopt}{_\mathrm{opt}}
\newcommand{\subOP}{_\mathrm{OP}}
\newcommand{\subat}{_\mathrm{at}}
\newcommand{\dc}{dc}
\newcommand{\rf}{rf}

\newcommand{\omegaL}{\omega_\mathrm{L}}

\newcommand{\nop}{\mathrm{o.p.}}
\renewcommand{\nop}{\boldsymbol{\mathcal{B}}_0}
\renewcommand{\nop}{\mathcal{B}_0}

\usepackage{siunitx}
\DeclareSIUnit\torr{Torr}
\DeclareSIUnit\amagat{amg}
\begin{document}
\newcommand{\thetitle}{Multiparameter quantum sensing and magnetic communications with a hybrid dc and rf optically pumped magnetometer}
\title[Article Title]{\thetitle}

\newcommand{\UW}{Centre for Quantum Optical Technologies, Centre of New Technologies, University of Warsaw, Banacha 2c, Warsaw, 02-097, Poland}
\newcommand{\ICFO}{ICFO - Institut de Ci\`encies Fot\`oniques, The Barcelona Institute of Science and Technology, 08860 Castelldefels (Barcelona), Spain}
\newcommand{\ICREA}{ICREA - Instituci\'{o} Catalana de Recerca i Estudis Avan{\c{c}}ats, 08010 Barcelona, Spain}
\newcommand{\fraunhofer}{Fraunhofer Centre for Applied Photonics, 99 George St, Glasgow G1 1RD, United Kingdom}

\author{Michał Lipka}
\email{mj.lipka@uw.edu.pl} 
\thanks{ML, AS contributed equally to this work.}
\affiliation{\UW}

\author{Aleksandra Sierant}
\email{aleksandra.sierant@icfo.eu}
%\thanks{Joint first authorship.}
\affiliation{\ICFO}

\author{Charikleia Troullinou}
\thanks{Present address: Fraunhofer CAP, Glasgow, UK.} 
\affiliation{\ICFO}
%\affiliation{\fraunhofer} 

\author{Morgan Mitchell}
\affiliation{\ICREA}
\affiliation{\ICFO}

\begin{abstract}
We introduce and demonstrate a hybrid optically pumped magnetometer (HOPM) that simultaneously measures one dc field  component and one rf field component quadrature  with a single atomic spin ensemble. The HOPM achieves sub-pT/$\sqrt{\mathrm{Hz}}$ sensitivity for both dc and rf fields, and is limited in sensitivity by spin projection noise  at low frequencies and by photon shot noise at high frequencies. We demonstrate with the HOPM a new application of multiparameter quantum sensing: background-cancelling spread spectrum magnetic communication. We encode a digital message as  rf amplitude, spread among sixteen 
channels from \SI{29}{\kilo\hertz} to \SI{33}{\kilo\hertz} in a noisy magnetic environment, and observe quantum-noise-limited rf magnetic signal recovery enabled by quantum-noise-limited dc noise cancellation, reaching noise rejection of \SI{15}{\decibel} at \SI{100}{\hertz} and more than \SI{20}{\decibel} at \SI{60}{\hertz} and below. We measure signal fidelity versus signal strength and extrinsic noise in communication of a short text message. The combination of high sensitivity, quantum-noise-limited performance, and real-world application potential makes the HOPM ideally suited for study of high-performance multiparameter quantum sensing.
\end{abstract}

\maketitle

Quantum sensing employs quantum systems and quantum measurements to acquire precise information about the sensed environment, and is formalized via the theory of quantum parameter estimation \cite{Helstrom1969}.
Quantum sensing of single parameters, e.g., phase shifts or magnetic fields, has been extensively studied both theoretically and experimentally \cite{Degen2017, PezzeRMP2018, BraunRMP2018}. Standard quantum limits (SQLs) to sensitivity and other metrics \cite{CavesPRD1981, HuelgaPRL1997} have been identified, and methods to surpass SQLs, known as quantum enhancement, have been devised \cite{CavesPRD1981, BenedictoPRL2022}. Proof-of-principle quantum enhancement has been demonstrated in optical interferometry \cite{GrangierPRL1987}, atomic clocks \cite{LerouxPRL2010, HostenN2016, HuangThesis2019} and optically pumped magnetometers (OPMs) \cite{WolfgrammPRL2010, Wasilewski2010, HorromPRA2012, Sewell2012}. Beyond-proof-of-principle quantum enhancement, in high-sensitivity practical instruments, has been demonstrated in a few cases \cite{Aasi2013NP, TsePRL2019Short, AcernesePRL2020, Troullinou2021, ZhengPRL2023}. These efforts revealed new aspects of quantum sensing, for example the role of measurement back-action in sensors operating beyond the shot-noise limit \cite{McCullerPRL2020, McCuller2021PRD, Troullinou2021}. 

Multiparameter quantum sensing (MPQS), the simultaneous estimation of two or more parameters with a quantum sensor \cite{SzczykulskaAPX2016, Demkowicz2020}, is a relatively new frontier for quantum parameter estimation, with potential application in many sensing tasks, e.g., measurement of intrinsically vector quantities such as fields, displacements or rotations. Recent work in MPQS \cite{ProctorPRL2018} has produced striking theoretical observations \cite{RehacekPRA2017, ProctorPRL2018, CarolloJSMTE2019, LiuJPAMT2020, HuangPRXQ2021, KaubrueggerPRXQ2023}, including the possibility of better sensitivity when estimating multiple parameters than when estimating parameters separately \cite{Gorecki2022} and proof-of-principle demonstrations \cite{ColangeloN2017, MollerN2017, PolinoO2019}.

Here we demonstrate a new experimental system for MPQS, a hybrid dc/rf optically pumped magnetometer (HOPM) that simultaneously estimates two magnetic field parameters from measurements on a single atomic spin ensemble. The HOPM has the potential for simultaneous quantum enhancement using optical \cite{Troullinou2021,bai2021quantum} and/or spin squeezing \cite{Sewell2012, Colangelo2017}, or by more exotic methods such as N00N states \cite{MitchellN2004, WolfgrammNP2013}. With this system, we demonstrate simultaneous, quantum-noise-limited \cite{TroullinouThesis2022}, sub-pT/Hz\textsuperscript{1/2} sensitivity in both parameters. Our HOPM implementation, based on alkali-vapor sensing methods, is suitable for miniaturization and manufacturing using proven techniques \cite{KitchingAPR2018}. Together, these features will enable use of MPQS methods in practical sensors.

\begin{figure}[t]
\centering
\includegraphics[width=\columnwidth]{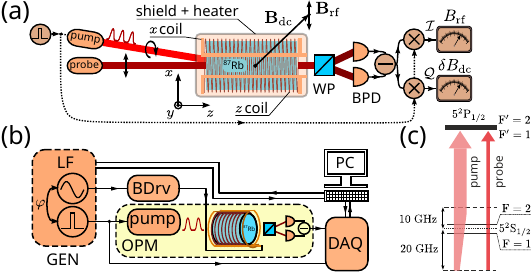}
 \caption{\textbf{Experimental setup of a hybrid \dc/\rf{} optically pumped magnetometer (HOPM).}
(a) HOPM consists of a shielded  $^{87}\mathrm{Rb}$ cell optically pumped (pump) in the Bell-Bloom configuration. Polarization rotation of the probe beam, proportional to the spin component $F_z$, is captured via a Wollaston prism (WP) and a balanced photodetector (BPD).  \dc~magnetic field $\bvdc{}$ in the $x$-$z$ plane at $45^\circ$ to the pump-probe direction allows hybrid \dc/\rf{} sensitivity.  \rfsym{} magnetic field $\bvrf{}$ is aligned along $x$ axis. (b) Configuration of the HOPM for sensitivity measurements and characterization of the $\isym{}/\qsym{}$ signals as functions of \dc/\rfsym{} magnetic fields. DAQ - data acquisition card,  BDrv - low noise current driver,  
%\gbbtext{(c) Measured quadratures $\isym{}$ and $\qsym{}$ versus  \dc{} (given by offset from the bias field $\delta\bdc{}$) and \rfsym{} magnetic fields (given by \rf{} field quadrature amplitude $\mathcal{X}$).  (d) Bell-Bloom pumping.  Probe beam is \SI{20}{\giga\hertz} blue-detuned from the $\mathrm{D}_1$ $F=1\rightarrow F'=2$ transition. The pump laser is \SI{20}{\giga\hertz} blue-detuned and rapidly swept to be \SI{10}{\giga\hertz} red-detuned once per modulation cycle.}  
(c) Bell-Bloom pumping.  Probe beam is \SI{20}{\giga\hertz} blue-detuned from the $\mathrm{D}_1$ $F=1\rightarrow F'=2$ transition. The pump laser is \SI{20}{\giga\hertz} blue-detuned and rapidly swept to be \SI{10}{\giga\hertz} red-detuned once per modulation cycle.}
\label{fig:idea_and_setup}
\end{figure}

A natural application of the HOPM is reception of ultra-low frequency (ULF), very-low frequency (VLF), and low frequency (LF) magnetic signals, employed in radio communications through weakly conducting media such as water \cite{Deans2018} or rock \cite{Gerginov2017}. In such media, the ULF/VLF/LF  bands benefit from lower attenuation than do higher frequencies, while also maintaining useful bandwidths. Potential applications include communications \cite{darpa_2016, page2021compact}, geotechnical \cite{gibson2010cave}, scientific  \cite{cohen_sensitive_2010} and  extraterrestrial exploration \cite{Fan2022}.

Due to the strong attenuation with distance in conducting media, both signal and environmental noise can be weak at the point of detection, leaving the receiver's intrinsic noise as the limiting factor.Using traditional antenna-based methods, a high sensitivity can be achieved at the cost of a large collection area, e.g., \SI{2}{\femto\tesla\per\sqrt\hertz} at \SI{30}{\kilo\hertz} with a \SI{1.69}{\meter\squared} antenna \cite{cohen_sensitive_2010}.  OPMs can be far more sensitive than ULF/VLF/LF antennas of similar size \cite{SavukovJMR2007}, allowing OPMs to function as ultra-compact radio receivers \cite{Ingleby2020}. OPM miniaturization can also greatly reduce sensor weight, important for carried or aerial use \cite{FunakiPS2014,becken2022semi}.  Together with piezoelectric-based direct antenna modulation techniques, which make possible \SI{10}{\centi\meter}-scale ULF/VLF/LF  transmitters \cite{Kemp2019}, miniaturization of ULF/VLF/LF  receivers may enable small-footprint duplex communication links.

Radio-frequency OPMs employ magnetic resonances that are tuned by the application of a dc bias field. This tuning requires compensation of changes in the ambient dc field, which can experience large changes as a craft reorients in the Earth field or other strong local field.  In the HOPM, such changes of the dc field can be monitored by the sensor itself, and feedback can be applied to maintain the desired tuning of the rf reception. When used in a field-locked-loop, this feedback allows both for cancellation of slow changes in the dc field and for selection of rf reception frequency. The simultaneous measurement of \rf{} and \dc{} fields is thus a MPQS problem that arises naturally in the application of OPMs to magnetic communications.

The article is organized as follows:  Section~\ref{sec:OPMDynamics} describes a Bloch-equation model for the spin dynamics and optical signal generation, applicable to a variety of OPM protocols, and reviews quantum enhancement in paradigmatic dc and rf OPM strategies. Section~\ref{sec:HOPMStrategy} introduces the hybrid dc/rf OPM strategy. Section~\ref{sec:OPMConstruction} describes the experimental implementation using an optically pumped \textsuperscript{87}Rb vapor. Section~\ref{sec:ModelValidation} describes experimental validation of the model. Section~\ref{sec:BackgroundRejection} describes a field-locked loop feedback system to stabilize the dc field seen by the atoms at a programmable value.  Section~\ref{sec:Sensitivity} describes sensitivity measurements, showing quantum-noise-limited performance. Finally, Section~\ref{sec:MagneticCommunication} describes VLF/LF magnetic communications using the HOPM. Section~\ref{sec:DiscussionAndOutlook} describes several natural extensions of the technique.

\begin{figure}[t]
 \includegraphics[width=\columnwidth]{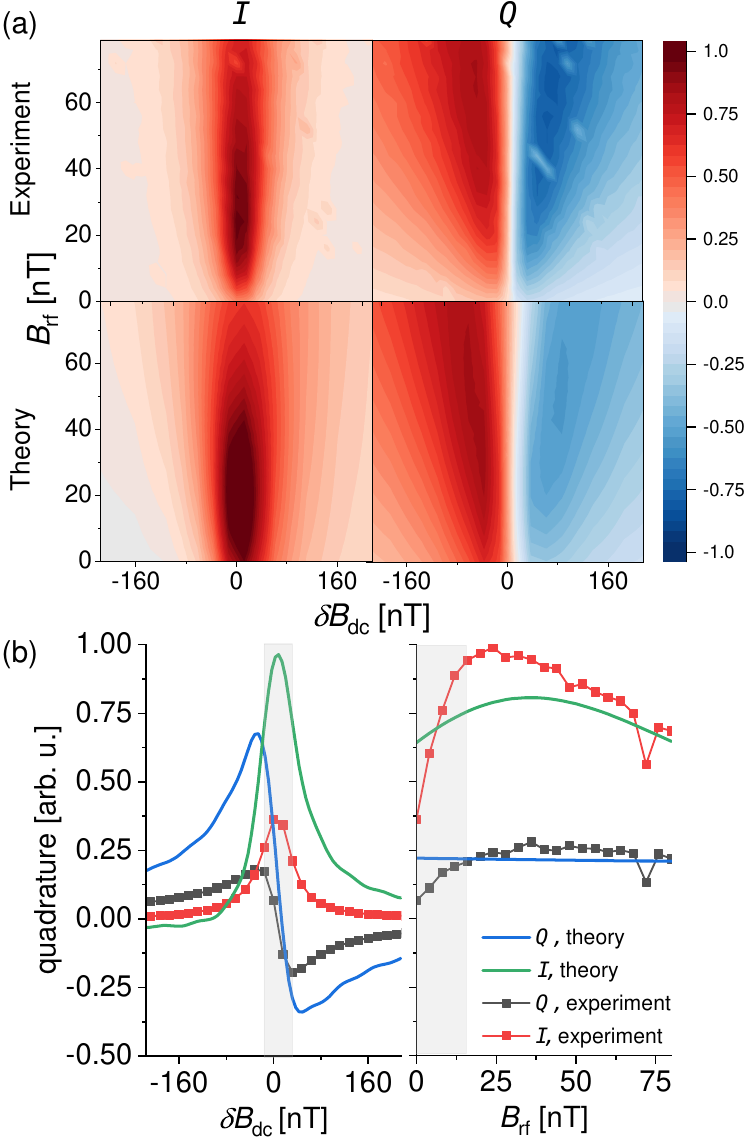}
    \caption{\textbf{Characterization of the HOPM.} 
   (a) Experimental and simulated dependence of the $\isym{}$ and $\qsym{}$ quadratures (rescaled by the same factor to achieve $\max_{\bdc,\brf} (\isym{},\qsym{}) =1$) 
    , for small changes $\delta B_\mathrm{\dc}$ in the offset magnetic field magnitude and for a range of small $\brf$ amplitudes with pump-resonant \rfsym{} field ($\Omega_\mathrm{\rfsym{}}=\Omega_p$, phase-locked).  Simulations by numerical integration of Eq.~(\ref{eq:SpinDynamics}), omitting the noise term. (b) Measured quadratures $\isym{}$ and $\qsym{}$ versus offset $\delta\bdc$ from resonance and $\brf$. Shaded bands indicate linear response region in which HOPM operates.
    %\ML{in (b) the resonance is shifted from $\delta B_{dc}=0$ (the band is not centered around 0) but $\delta B_{dc}$ is defined to be 0 at resonance. It also makes the right subplots cross section to be at a slightly wrong point. Maybe we should shift the $\delta B_{dc}$ coordinate? Or we can just treat as an experimental error I guess}
    }
    \label{fig:results}
\end{figure}

\section{OPM dynamics and single-parameter estimation strategies }
\label{sec:OPMDynamics}
In an optically pumped alkali vapor, a vector collective spin $\mathbf{F}$ evolves by a stochastic differential equation containing the magnetic field as a parameter, for example the  Bloch equation 
\begin{eqnarray}
\label{eq:SpinDynamics}
\frac{d}{dt}\bF(t)  &=&  \left[-\gamma \mathbf{B}(t) + G_S S_3(t) \hat{z} \right] \times \bF(t)  - \Gamma  \bF(t)   
\nonumber \\ & & + R\subOP(t) [\bFmax -\bF(t)] +  \bN\subat(t)
\end{eqnarray}
where $\gamma$ is the atomic gyromagnetic ratio, %of $^{87}$Rb, 
$\mathbf{B}$ is the magnetic field vector, $\Gamma$ is the spin relaxation rate, $R\subOP$ is the optical pumping rate, $\bFmax$, proportional to the number of atoms in the ensemble, is the value $\bF$ would take if fully-polarized along the optical pumping direction, $G_S S_3(t) \hat{z}$ describes optical Zeeman shifts causing measurement back-action and $\bN\subat$ is a Langevin noise term describing spin fluctuations \cite{Troullinou2021}.  The spin evolution is read out by off-resonance Faraday rotation, described by the input-output relation \cite{KongNC2020}
\begin{eqnarray}
\label{eq:S2Out}
S_2^{\mathrm{(out)}}(t)&=&  S_1^{\mathrm{(in)}}(t)\sin{\phi}(t) + S_2^{\mathrm{(in)}}\cos{\phi}(t) 
\nonumber \\ & \approx &
G_F  F_z(t) S_1\supin(t) + N\subopt(t),
\end{eqnarray}
where $S_\alpha$, $\alpha \in \{1,2,3\}$ indicate Stokes parameters before $(\supin)$ or after $(\supout)$ the atoms, $\phi = G_F F_z$  is the Poincare-sphere rotation angle, $G_F$ is a detuning-dependent coupling factor, $\hat{z}$ is the propagation direction of the probe light, and $N\subopt$ is the quantum polarization noise of the detected Stokes component \cite{PhysRevD.23.1693}. The approximation holds for small $\phi$. 

Two paradigmatic single-parameter quantum sensing approaches have been studied both theoretically and experimentally with OPMs: A canonical scalar magnetometer estimates $|\mathbf{B}|$ by orienting $\bB$ orthogonal to the readout direction ($z$), controlling $R\subOP$ to generate polarization orthogonal to $\bB$, and observing the Larmor frequency $\omegaL \equiv \gamma |\bB|$ with which $\bF$ precesses about $\bB$.  It has been shown that this technique is naturally back-action evading \cite{ColangeloN2017, ColangeloPRL2017}, and single-parameter sensitivity enhancement has been demonstrated \cite{Troullinou2021}.  A canonical radio-frequency (rf) magnetometer estimates a field $\bB(t) = \bB_\mathrm{dc} + \bB_\mathrm{rf}(t)$, where $\bB_\mathrm{rf}(t)$ is relatively weak, sinusoidally varying, and orthogonal to the constant and known $\bB_\mathrm{dc}$, which is perpendicular to the readout direction ($z$). A phase-sensitive measurement of $\bB_\mathrm{rf}(t)$ is performed by controlling $R\subOP$  to generate polarization along $\bB_\mathrm{dc}$. In this strategy $\bF$ has small $F_z$ component unless resonantly driven, i.e. if $\bF_\mathrm{rf}(t)$ has components oscillating at or near $\omegaL$. In effect, the readout detects the earliest stages of the magnetic Rabi oscillation.  With continuous probing, this allows estimation of both quadratures of $\bB_\mathrm{rf}(t)$  but does not evade measurement back-action, a condition that restricts the sensitivity to the SQL. With stroboscopic probing, i.e., a probe power $S_1(t)$  consisting of pulses spaced by an integer number of half-cycles, back-action is evaded, but only one quadrature can be estimated \cite{Shah2010}. Spin squeezing \cite{Vasilakis2015, KongNC2020, bao2020} noise squeezing \cite{guarrera2019parametric, guarrera2021spin} and single-parameter sensitivity enhancement \cite{MartinCiurana2017, ZhengPRL2023} have been demonstrated with such stroboscopic techniques.

The HOPM we present below combines a scalar and rf magnetometer in a single measurement protocol, with the possibility of back-action evasion and sensitivity enhancement when estimating both parameters simultaneously. As in the scalar OPM, the spin precession frequency is used to infer $|\bB|$, and as in the rf OPM, the amplitude of the spin precession signal is used to infer one quadrature of the rf field. 
%This is made possible by a skew optical pumping/probing scheme, in which one of the trio $\{\bB_\mathrm{dc},\bFmax, z\}$ is at an intermediate angle (neither parallel nor perpendicular) with respect to the others. With this arrangement, optical pumping creates a polarization whose amplitude indicates one quadrature of $\bB_\mathrm{rf}(t)$, while its frequency indicates $\bB_\mathrm{dc}$. 
A natural extension, which we also implement, is to operate with closed-loop control of $\bB_\mathrm{dc}$, to tune $\omegaL$ so as to maintain resonance in a fluctuating magnetic environment. 

% \section{Experimental system}

\newcommand{\supnom}{^{(\mathrm{nom})}}
 
\section{hybrid-OPM multiparameter estimation strategy}
\label{sec:HOPMStrategy}
The setup and operation of the hybrid \dc/\rf{} OPM are illustrated schematically in Fig. \ref{fig:idea_and_setup}. Details of the apparatus are given in Section~\ref{sec:OPMConstruction} and in \cite{Troullinou2021}.  A \dc{} field $\bvdc$ is applied, nominally along the $(\hat{x}+\hat{z})/\sqrt{2}$ direction and with magnitude $\bdc$.
As in Bell-Bloom (BB) magnetometry \cite{bell1961optically}, an optical pumping beam along $\hat{z}$ drives the atomic spin $\bF$ with a pumping rate $R_\mathrm{OP}(t)$, a periodic function of $t$ with period $2\pi/\Omega_p$. When the optical pumping frequency approaches the Larmor frequency, i.e., when $\delta\bdc \equiv \bdc - \Omega_p/\gamma$ is small, it produces a resonant build-up of spin polarization, in which $F_z$ and thus $\phi$ oscillate at frequency $\Omega_p$. The rotation signal $S_2\supout(t)$ is demodulated with a digital lock-in amplifier phase-referenced to $R_\mathrm{OP}(t)$, to obtain $\isym{}$ and $\qsym{}$, the in-phase and \SI{90}{\degree} quadrature components, respectively. The demodulation phase is chosen such that $\qsym{} = 0$ at resonance, i.e., for $\delta\bdc = 0$.

An \rf{} field $\bvrf{}(t) =   [\mathcal{X}(t) \cos \Omega_\mathrm{\rfsym{}} t + \mathcal{Y}(t) \sin \Omega_\mathrm{\rfsym{}} t] \hat{x}$, %in the $\hat{x}-\hat{z}$ plane, 
with quadrature amplitudes $\mathcal{X}(t)$, $\mathcal{Y}(t)$ drives a magnetic resonance if the oscillation of $\bvrf{}(t)$ contains components near $\omega_L$. Without loss of generality, we take the carrier frequency $\Omega_\mathrm{\rfsym{}}$ equal to $\Omega_p$ with phase lag $\varphi$, so that $\mathcal{X} = B_\mathrm{rf} \cos\varphi$, $\mathcal{Y} = B_\mathrm{rf} \sin\varphi$, where $B_\mathrm{rf}$ is the amplitude of the \rf~ drive. Measured dependence of  $(\mathcal{I},\mathcal{Q})$ on $\varphi$ is discussed in Appendix \ref{app:varphi}. For  $\varphi=n\pi, n\in \mathbbm{Z}$, we have $\mathcal{Y}=0$, hence $\mathcal{X}$ encodes the amplitude of the \rf{} field $B_\mathrm{rf}$, which is our operating parameter of interest and will be used in the manuscript from now on.
This drive modifies the $\mathbf{F}$ oscillation and is reflected in the $\isym{}$ and $\qsym{}$ signals. This is shown in Fig.~\ref{fig:results} (a), which compares the response to $\bdc$ and $\isym{}$, as predicted by numerical integration of Eq.~(\ref{eq:SpinDynamics}), omitting stochastic terms and by direct measurement. Relevant cross sections of the Fig.~\ref{fig:results} (a) results are shown in Fig.~\ref{fig:results} (b). From these, we observe that
$\qsym{}$, which near resonance encodes the phase of the signal relative to the drive, principally responds to $\bdc$, while $\isym{}$, which near resonance encodes the amplitude of the signal, principally responds to $B_\mathrm{rf}$.
That is, 
\begin{eqnarray}
     \left| \frac{\partial \mathcal{I}}{\partial B_\mathrm{rf}} \right|_{\nop} 
\gg
 \left| \frac{\partial \mathcal{I}}{\partial\bdc} \right|_{\nop}
    \label{eq:iquad}
&\mathrm{and} &     
     \left| \frac{\partial \mathcal{Q}}{\partial\bdc} \right|_{\nop}
\gg
\left| \frac{\partial \mathcal{Q}}{\partial B_\mathrm{rf}} \right|_{\nop},
\end{eqnarray}
where $\nop$ indicates the nominal operating point $\delta\bdc = B_\mathrm{rf} = 0$. The above relations hold also in the neighborhood of $\nop$, in the ranges indicated by shaded bands in Fig.~\ref{fig:results} (b). Relative to experiment, theory overestimates the dc response $d\mathcal{Q}/d\bdc$ and underestimates the rf response $d\mathcal{I}/d\brf$. This may due to the simplicity of the spin dynamics model, Eq.~( \ref{eq:SpinDynamics}), which describes only one of two hyperfine states, and greatly simplifies the optical pumping process \cite{Troullinou2021, MouloudakisPRA2022}. Qualitative agreement seen in \autoref{fig:results} nonetheless suggests the model correctly captures the essentials of HOPM operation.

\section{HOPM construction and operation}
\label{sec:OPMConstruction}

The experimental setup of the HOPM is shown in Fig.~\ref{fig:idea_and_setup}~(a)-(b).
Isotopically enriched \textsuperscript{87}Rb and \SI{100}{\torr{}} of N$_2$ buffer gas is contained in a cell with \SI{3}{\centi\meter} internal path, placed inside a ceramic oven. A temperature of \SI{105}{\celsius} is maintained by intermittent Joule heating, producing a \textsuperscript{87}Rb vapor density of \SI{8.2e12}{atoms\per\centi\meter\cubed}. Surrounding induction coils in the $x$ and $z$ direction are driven by a low-noise current supply (TwinLeaf CSUA300 and either TwinLeaf CSUA300 or Koheron DRV300‑A‑40,
% \ML{check model}
respectively), generating the offset \dc~field of  $\bdc{}\approx \SI{4.3}{\micro\tesla}$, corresponding to $\omegaL\approx\SI{32}{\kilo\hertz}$, and generating the rf field.  Four layers of mu-metal shielding ensure environmental magnetic isolation.
The OPM is pumped by a circularly polarized beam from a distributed Bragg reflector laser with power \SI{200}{\micro\watt} unless otherwise specified, current modulated  at frequency $\Omega_p$ (Sigilent SDG1025) in a Bell-Bloom scheme to generate a periodic pumping rate $R_\mathrm{OP}(t)$.  The collective atomic spin precession in the \dc/\rf{} magnetic field is observed with a shot-noise-limited polarimeter (Thorlabs PDB450A after polarization-splitting optics) sensing the polarization rotation of a linearly polarized \SI{500}{\micro\watt}, \SI{20}{\giga\hertz} blue-detuned probe beam from a tunable frequency-doubled diode laser with a tapered amplifier (Toptica TA-SHG 110) or interchangeably from an external cavity diode laser (Toptica DL100). Parameters of pump (power, detuning, duty cycle) and probe (power, detuning) are optimized as in \cite{troullinou2021squeezed} for achieving the highest sensitivity for both quadratures, giving priority to DC the field. An \rfsym{} signal from a function generator drives induction coils through a low-noise controller to generate $\bvrf{}$. The signal driving the pump laser's current is phase ($\varphi$) and frequency synchronized with the rf oscillations.  A DAQ records signals from the BPD and the pump current monitor.

\begin{figure}[t]
    \includegraphics[width=\columnwidth]{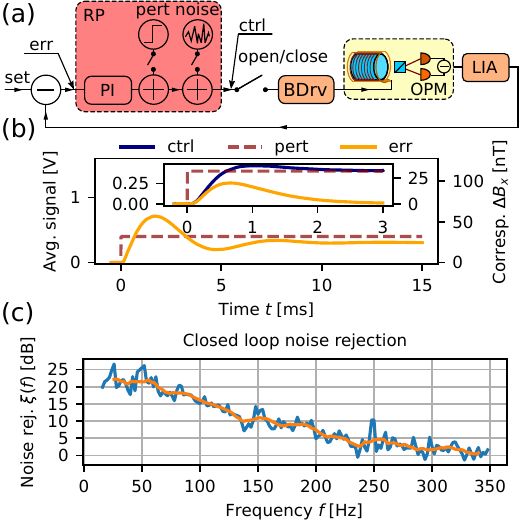}
    \caption{\textbf{Feedback loop characterization.} (a) External magnetic field stabilization system. Lock-in amplifier's (LIA) out-phase quadrature $\qsym{}$  provides feedback signal proportional to the deviation of the Larmor frequency from the optically pumped magnetometer's (OPM) pumping frequency. Redpitaya (RP) board implements a proportional-integral controller (PI) and addition of a test perturbation signal (pert) in the digital domain. Control signal (ctrl) alters the %\btext{\sout{magnetic field} 
    $B_z$ magnetic field component via a current controller (BDrv) driving induction coils.  (b) Averaged (10 repetitions) open loop (main plot) and close loop (inset) responses to a ca. \SI{32}{\nano\tesla} step perturbation of
    % 0.4 V x 1 mA / V x 80 nT / mA
    the $B_x$ magnetic field component. (c) Noise rejection $\xi(f)$ (blue) and $\xi(f)$ smoothed with a \SI{20}{\hertz} moving average (orange), see text. 
    }
    \label{fig:loopresp}
\end{figure}

\section{Model validation and signal characterization}  
\label{sec:ModelValidation}
Fig. \ref{fig:results} shows the dynamical characterization of the HOPM, using experimental setup presented in Fig.~\ref{fig:idea_and_setup}~(b). 
For $\isym{}/\qsym{}$ characterization a single frequency \rfsym{} carrier, phase-locked ($\varphi=0$) with the pumping signal, is employed with a range of amplitudes. $\isym{}$ and $\qsym{}$ are recorded as a function of applied $\brf{}$ and $\delta\bdc{}$. Results are shown in Fig.~\ref{fig:results} and show a good agreement with model predictions.

\section{Field-locked-loop and dc background rejection}
\label{sec:BackgroundRejection}

We implement a field-locked loop, i.e., feedback from the measured $\mathcal{Q}$ to the applied $\delta\bvdc$, with set-point $\mathcal{Q} = 0$. This maintains the $\mathcal{I}$ responsivity $c_\mathcal{X}$ near its maximum value in the presence of external perturbations to $\bvdc$. The implemented servo-loop is shown in Fig.~\ref{fig:loopresp} (a). A digital-domain proportional-integral controller, implemented with an FPGA board (RedPitaya STEMlab 125-14 running PyRPL), drives an induction coil controller (TwinLeaf CSUA300) feeding the $B_z$ coil. In this way, we obtain a fast closed-loop response and high noise rejection, as shown in Fig. \ref{fig:loopresp} (b), (c), respectively. 

To quantify the noise rejection, we operate the HOPM both in open-loop (OL), i.e., without feedback to $\bvdc$, and in closed-loop (CL), i.e., with feedback, both with added noise and without, and observe the resulting PSD $\mathcal{P}$ of the demodulated signal $\mathcal{I}$. The added noise was 
$\approx \SI{10}{\pico\tesla\per\sqrt\hertz}$ white noise, filtered with successive \SI{1}{\kilo\hertz} low-pass and \SI{1}{\hertz} high-pass first-order digital filters and added in the digital domain, as depicted in Fig. \ref{fig:loopresp} (a). The resulting  noise rejection factor
\begin{equation}
\xi(f) \equiv 
\frac
{
\mathcal{P}^{\mathrm{OL}}_{\mathrm{n}}(f) - \mathcal{P}^{\mathrm{OL}}_{\mathrm{c}}(f)
}
{
\mathcal{P}^{\mathrm{CL}}_{\mathrm{n}}(f) - \mathcal{P}^{\mathrm{CL}}_{\mathrm{c}}(f)
}
\end{equation} 
is shown in Fig.~\ref{fig:loopresp} (c). We observe a response time of \SI{3}{\milli\second} and thus a few-hundred \SI{}{\hertz} bandwidth, and a measured background rejection of more than \SI{20}{\decibel} below \SI{60}{\hertz}, dropping to \SI{3}{\decibel} at about \SI{260}{\hertz}. We note that feedback to a single component, here $B_z$, is sufficient to control $\bdc$, and thus the \rf{} reception frequency, regardless of the direction of the \dc{} magnetic perturbation.

\begin{figure}
    \includegraphics[width=\columnwidth]{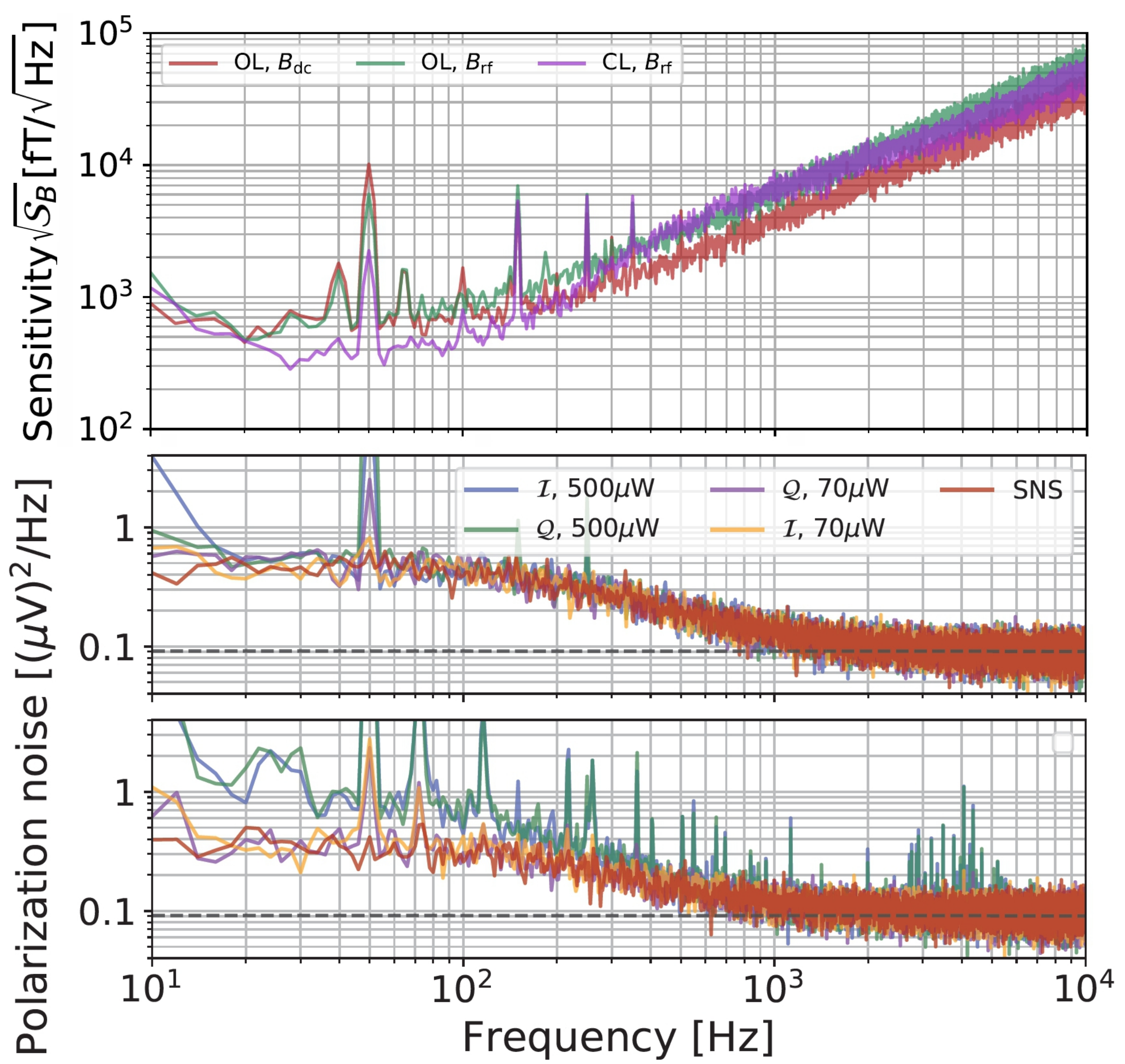}
    \caption{\textbf{Sensitivity and polarization noise after demodulation.} Top: sensitivity of the magnetometer to a small harmonic perturbation at frequency $f$ of: (OL, $B_\mathrm{\dc}$) the offset magnetic field magnitude $\bdc{}$ with open feedback loop; (OL/CL, $\brf$) amplitude of the resonant ($\Omega_\mathrm{\rfsym{}}=\Omega_p$) \rfsym{} magnetic field with open (OL) or closed (CL) feedback loop. Bottom: photon shot noise (PSN) and spin projection noise (SPN), and measurement back-action (MBA) in the HOPM. Upper graph shows, for reference, the Faraday rotation spectrum in single-parameter \dc{} field estimation, i.e., with the \dc{} field along $\hat{x}$, as in \cite{Troullinou2021}. Red curve shows power spectrum in the absence of optical pumping (spin noise spectroscopy, SNS), other solid curves show power spectra for the $\isym$ and $\qsym$ quadratures at different optical pumping levels, as indicated in the plot legend. Dashed line shows the inferred photon shot noise level. The observed quantitative agreement of spectra for the OPM and for SNS (which is intrinsically insensitive to  technical noise affecting the spins), indicates the absence of other noise sources. 
    Lower graph shows the corresponding spectra for the HOPM, i.e., with the \dc{} field along $(\hat{x} + \hat{z})/\sqrt{2}$. Apart from noise spikes, the spectra show agreement of SNS and HOPM noise levels for weak, \SI{70}{\micro\watt} optical pumping, indicating that the spectra are dominated by a combination of PSN and SPN. In contrast, for strong, \SI{500}{\micro\watt} optical pumping, the system becomes sensitive enough in the \SI{20}{\hertz} to \SI{200}{\hertz} band to reveal low-frequency magnetic noise and MBA, which from stochastic simulation of \autoref{eq:SpinDynamics} has the same spectral shape as the SPN but grows with spin polarization and thus with optical pumping.
    }
    \label{fig:QuantumNoiseLimited}
\end{figure}

\section{Quantum-noise-limited sensitivity} 
\label{sec:Sensitivity} 
We write the power spectral density (PSD) for an observed quantity $\alpha$ as $\mathcal{S}_\alpha(f)$, where $f$ is the linear frequency. We also use this notation for the \dc{} and \rf{} magnetic sensitivities, i.e., we write the equivalent magnetic noise $\mathcal{S}_\alpha(f),\;\alpha\in\lbrace\bdc{},\brf\rbrace$. Sensitivity spectra of the HOPM were measured as in prior work with BB magnetometers \cite{Gerginov2020, Gerginov2017, Jimenez-Martinez2010}: Coils and current drivers are calibrated by measurement of Larmor frequency in a low-density atomic vapor. The HOPM is then operated as described above, near its nominal operating point $\delta \bdc = 0$ and $\brf = 0$.  The response, defined as $R_\mathcal{I}(f) \equiv d \mathcal{I}(f)/ d\brf(f)$ ($R_\mathcal{Q}(f) \equiv d \mathcal{Q}(f)/ d \bdc(f)$) is measured in two steps. First, we acquire quasistatically 
$\mathcal{I}$ versus $\brf$  with  $\delta\bdc = 0$ ($\mathcal{Q}$ versus $\bdc$ with $\brf = 0$) and make a linear fit around the operating point to find $R_\mathcal{I}(0)$ ($R_\mathcal{Q}(0)$). Second, for a range of frequencies $f$, the ratio $R^2_\mathcal{I}(f)/R^2_\mathcal{I}(0) = S_\mathcal{I}(f)/S_\mathcal{I}(0)$ ($R^2_\mathcal{Q}(f)/R^2_\mathcal{Q}(0) = S_\mathcal{Q}(f)/S_\mathcal{Q}(0)$) is measured by applying (with the function generator labelled LF in Fig. \ref{fig:idea_and_setup}b) a small harmonic perturbation to $\brf$ ($\bdc$), i.e., at frequency $f$ and within the linear response regime, and recording $S_\mathcal{I}(f)$ ($S_\mathcal{Q}(f)$), obtained by Fourier transform with a Hann window. This ratio method  automatically accounts for frequency dependence of the signal chain and data analysis. The HOPM is then operated at the nominal operating point with no applied signal, and the residual noise PSD $S_\mathcal{I}(f)$ ($S_\mathcal{Q}(f)$) is recorded in the same way. The sensitivity is then calculated as $S_\mathcal{\brf}(f) = S_\mathcal{I}(f) R_\mathcal{I}^{-2}(f)$ ($S_{\bdc}(f) = S_\mathcal{Q}(f) R_\mathcal{Q}^{-2}(f)$).
%\gtext{cnt: shouldn't the following be changed to $S_{\brf}(f)$ to be consistent with the change of symbols earlier? }

Figure \ref{fig:QuantumNoiseLimited} shows measured sensitivities in closed-loop (CL) operation, stabilizing $\qsym{}=0$ and thus $\delta\bdc = 0$, and in open-loop (OL) operation, with $\delta\bdc$ set to zero at the start of the acquisition and limited by passive stability.  The experimental setup of the HOPM, configured for sensitivity measurements, is shown in Fig. \ref{fig:idea_and_setup} (b). Apart from a few noise spikes, e.g., at multiples of the \SI{50}{\hertz} power-line frequency, and ``1/f noise'' below \SI{20}{\hertz}, the sensitivities, which are sub-\SI{}{\pico\tesla\per\sqrt\hertz} in the \SI{10}{\hertz}-\SI{200}{\hertz} band of interest, are quantum noise limited, as we now show, using the methodology of Troullinou \textit{et al.} \cite{Troullinou2021}. %\vtext{as:cnt: need to specify again which spectrum we are talking about, I could see the quantum noise limited performance in the upper fig4 but I do not know if it so straightforward to infer in the lower picture spectrum, unless I am missing something, also the model in the 2021PRL paper refers to the case of a single component Bdc in the perpendicular direction so that the AC stark shift affects the out of plane F components here that   }

Operating the system as in the sensitivity measurement just described, but at negligible atomic vapor density (achieved by allowing the cell to cool to room temperature), we observe linear scaling of $S_\mathcal{I}$ and $S_\mathcal{Q}$ with probe power, confirming photon-shot-noise (PSN)-limited probing and establishing the PSN level. This same noise level is observed in HOPM operation in the frequency regime above \SI{1500}{\hertz}, see Fig.~\ref{fig:QuantumNoiseLimited}, confirming PSN-limited operation in this regime. Using spin-noise spectroscopy (signal acquisition in the presence of atoms and probe light but without optical pumping), we observe that the spectral region from \SI{20}{\hertz} to \SI{200}{\hertz} is dominated by spin projection noise (SPN), which exceeds PSN in this frequency regime. For the OPM in CL operation, we find the same noise level in this range, apart from a noise spike at the mains frequency. This confirms that this low-frequency regime is SPN-dominated. For intermediate frequencies, the noise is dominated by a mixture of SPN and PSN.

\begin{figure*}[t]
    \includegraphics[width=0.9\textwidth]{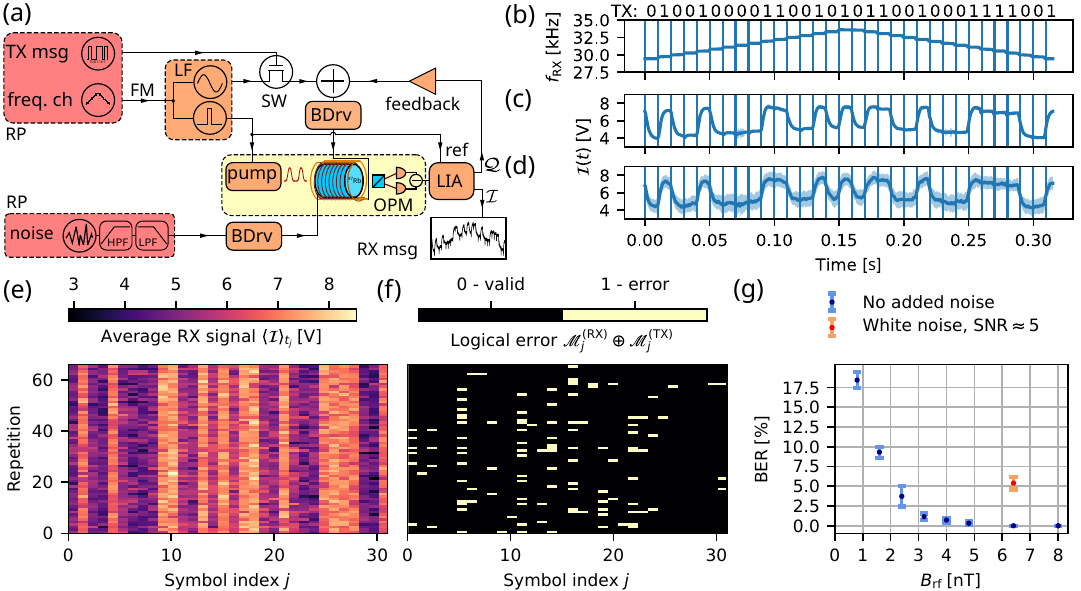}
    \caption{\textbf{Receiver operation.} (a) OPM experimental setup configured as a magnetic \rfsym{} receiver. Frequency modulation (FM) of \rfsym{} carrier and OPM pumping signals selects the frequency channel $f_\mathrm{RX}^{(j)}$ for each $j$-th symbol. Radio-frequency switch (SW) allows on-off-keying (OOK) of the \rfsym{} carrier according to the binary message (TX msg, $\mathscr{M}^{\mathrm{(TX)}}$). A feedback loop  on the $\qsym{}$ quadrature corrects the offset magnetic field $B_x$ component for the Larmor frequency to match $f_\mathrm{RX}$. Amplitude variation of the $\isym{}$ quadrature reveals the received symbols (RX msg, $\mathscr{M}^{\mathrm{(RX)}}$). Additional white Gaussian noise is introduced into the $z$-axis magnetic field $B_z$ to simulate unshielded operation. (b) Carrier frequency $f_\mathrm{RX}$ switches symbol-to-symbol. (c),(d) Exemplary receiver signals with $\brf{}=\SI{6.39}{\nano\tesla}$ 
    %\mtext{mwm: this was $\brf{}=\SI{4.52}{\nano\tesla}$ RMS, but now $\brf$ is defined as the amplitude.  the x-axis of graph (e) should be changed to agree with this definition. } 
    for a 32-bit transmitted message (TX) with the shade corresponding to one standard deviation over $N_\mathrm{TX}=67$ repetitions. (c) Without additional noise. (d) With white noise, here and henceforth at the signal-to-noise ratio (SNR) of $\approx 5$. (e) Average bit error rate (BER) of the received message. 
    (f) Receiver signal $\isym{}$ averaged over $j$-th symbol window $t_j$ and digitized signal (g) $\mathscr{M}^{\mathrm{(RX)}}$ compared (bitwise XOR) with the TX message $\mathscr{M}^{\mathrm{(TX)}}$, with white noise. }
    %\bbbtext{AS: Michal, can we swap (e), (f), (g) info (f), (g), (e)?}}
    \label{fig:rx}
\end{figure*}

\section{VLF/LF magnetic communication with the \MakeLowercase{h}OPM}
\label{sec:MagneticCommunication}

To demonstrate background-cancelling magnetic communication, we configure the hybrid magnetometer as a frequency-hopping spread-spectrum (FHSS) magnetic field \rfsym{} receiver (RX), see Fig. \ref{fig:rx} (a). To generate a  
VLF/LF signal, a 32-bit message is encoded with on/off keying (OOK) of $\bvrf$ (generated with a coil inside the shielding) with amplitude $\brf{}=\SI{6.39}{\nano\tesla}$,  and symbol rate $f_\mathrm{OOK}=\SI{100}{\hertz}$. Transmitted symbols are spread over 16 frequency channels $f_\mathrm{RX}^{(1)}$ to $f_\mathrm{RX}^{(16)}$ with a separation of $\SI{250}{\hertz}$ and the channel hopping scheme $1\rightarrow2\rightarrow\ldots
\rightarrow16\rightarrow16\rightarrow15\rightarrow \ldots\rightarrow1$, as shown in Fig.~\ref{fig:rx}~(b). The RX reception frequency is tuned in the same sequence, with $\Omega_p = 2 \pi f_\mathrm{RX}$, and with $\bvdc$ set by feedback as in the closed-loop operation described above. Representative RX waveforms ($\isym{}(t)$ quadrature) in the presence and absence of external noise (see Appendix for details) are shown in Fig. \ref{fig:rx} (c), (d), together with the originally transmitted message (TX). 

As shown in Fig.~\ref{fig:rx}~(f) [also visible in  Fig.~\ref{fig:rx}~(d)] there are systematic deviations in the RX signal before digitization, with a simple thresholding strategy leading to symbol misclassification (details on threshold selection are given in Appendix). Logical errors are depicted in Fig. \ref{fig:rx} (g). A visual inspection of the signal quality in Fig.~\ref{fig:rx}~(c),(d) suggests that a tailored filtering and classification algorithm could greatly reduce the error rate.  
Fig. \ref{fig:rx}~(e) shows the measured bit error rate (BER) as a function of carrier amplitude, which shows the expected super-exponential scaling with signal strength. Also shown is the BER for $\brf = \SI{6.39}{\nano\tesla}$  with added noise, corresponding to Fig.~\ref{fig:rx}~(d),(f),(g). 
% \begin{equation}
%     \varepsilon=\langle \mathscr{M}_\mathrm{RX} \oplus \mathscr{M}_\mathrm{TX} \rangle_{N_\mathrm{TX}} 
% \end{equation}

\section{Discussion and outlook}
\label{sec:DiscussionAndOutlook}

Due to the role of the \dc~magnetic field in tuning the HOPM-based receiver, its operation is inevitably prone to \dc~magnetic noise. In particular, low frequency components within the HOPM rf bandwidth are most detrimental. We have demonstrated how a simple all-digital control system with a single proportional-integral (PI) controller reduces noise power spectral density in this frequency range up to about \SI{20}{\decibel}. In the time domain, we observed a stabilization of a rapid \SI{32}{\nano\tesla} perturbation of $\bdc{}$ in about \SI{3}{\milli\second}. We note that in the FHSS scheme the RX is synchronized with the TX. A straightforward improvement to our proof-of-principle demonstration would thus be to add a feed-forward magnetic field compensation, to make the Larmor frequency match the TX channel already from the moment it switches. Not only would this allow a faster settling time (hence higher TX/RX bandwidth) but also pseudo-random channel hopping and larger separations between frequency channels. A feedback loop would nonetheless remain an essential part of the system, to compensate the low frequency environmental noise present in any unshielded system. With the sub-\SI{}{\pico\tesla\per\hertz\tothe{1/2}} sensitivity of the HOPM, such a simple feedback loop can provide a highly precise control, compensating slow changes in background field of any strength.  The ability to follow sudden changes is in this implementation limited by the LIA bandwidth, which can be up to twice the carrier frequency.

For simplicity our TX/RX demonstration uses the on-off-keying (OOK) encoding which maps the bit values to a two-state amplitude modulation (AM). The HOPM is also compatible with frequency- and phase-modulation encodings, e.g., minimal shift keying (MSK) encoding, currently a standard in high-power ULF/VLF/LF systems
\cite{cohen_sensitive_2010}.

\section{Conclusions}
\label{sec:Conclusions}
We have described a hybrid optically pumped magnetometer (HOPM) that uses a single atomic ensemble to simultaneously measure dc and rf field components with quantum-noise-limited sub-pT/$\sqrt{\mathrm{Hz}}$ sensitivity.  A need for simultaneous dc/rf sensing arises naturally in an important emerging application of atomic sensors -  background-cancelling VLF/LF magnetic communication with ultra-compact receivers, which enables robust communication. The HOPM is thus a practical example of high-sensitivity multiparameter quantum sensing.

The high \dc~field sensitivity allows self-adaptation on few-\SI{}{\milli\second} time-scales to perturbations of the magnetic environment, rejecting magnetic field noise by more than \SI{20}{\decibel} at low frequencies and with a few \SI{100}{\hertz} bandwidth.  Using this capability, we have shown quantum-noise-limited reception of an on-off keyed signal spread over 16 frequency-hopping channels separated by \SI{250}{\hertz}, in the presence of externally introduced magnetic field noise, simulating unshielded operation. 

Simple modifications will allow more sophisticated protocols including quadrature coding, pseudo-random spread spectrum, and improved fidelity by model-aware signal processing. The technology shows the ability of multiparameter quantum sensing methods to meet application-specific sensor requirements, and opens new directions for quantum-enhanced atomic sensing and magnetic communications.

\section{Data availability}
Data for figures 2-6 has been deposited at \cite{RepositoryData}.

\section{Acknowledgments}
We would like to thank Michał Parniak, Sven Bodenstedt, Kostas Mouloudakis, and Vito Giovanni Lucivero for helpful readings of the manuscript.
This project has received funding from the European Defense Fund (EDF) under grant agreement EDF-2021-DIS-RDIS-ADEQUADE (n°101103417). Authors acknowledge support from the Government of Spain (Severo Ochoa CEX2019-000910-S), Fundació Cellex, Fundació Mir-Puig, and Generalitat de Catalunya (CERCA, AGAUR). This study was supported by the European Commission project OPMMEG (101099379), Spanish Ministry of Science MCIN and NextGenerationEU (PRTR-C17.I1) and by Generalitat de Catalunya ``Severo Ochoa'' Center of Excellence CEX2019-000910-S; projects  SAPONARIA (PID2021-123813NB-I00) and MARICHAS (PID2021-126059OA-I00) funded by MCIN/ AEI /10.13039/501100011033/ FEDER, EU; Generalitat de Catalunya through the CERCA program;  Ag\`{e}ncia de Gesti\'{o} d'Ajuts Universitaris i de Recerca Grants No. 2017-SGR-1354 and 2021-SGR-01453;  Secretaria d'Universitats i Recerca del Departament d'Empresa i Coneixement de la Generalitat de Catalunya, co-funded by the European Union Regional Development Fund within the ERDF Operational Program of Catalunya (project QuantumCat, ref. 001-P-001644); Fundaci\'{o} Privada Cellex; Fundaci\'{o} Mir-Puig;
Funding: Fundacja na rzecz Nauki Polskiej (MAB/2018/4 “Quantum Optical Technologies”); European Regional Development Fund; Narodowe Centrum Nauki (2021/41/N/ST2/02926).
The “Quantum Optical Technologies” project is
carried out within the International Research Agendas programme of the
Foundation for Polish Science co-financed by the European Union under the
European Regional Development Fund.
This research was funded in whole or in part by National Science Centre, Poland 2021/41/N/ST2/02926. ML was supported by the Foundation for Polish Science (FNP) via the START scholarship. ML was also supported by the 1st competition for co-financing the mobility of doctoral candidates at the University of Warsaw under Action IV. 4.1 ``A complex programme of support for UW PhD students'' funded by the ``Excellence Initiative – Research University (2020-2026)'' programme of the Ministry of Science and Higher Education, Poland.
~\\~\\
Funded by the European Union. Views and opinions expressed are however those of the author(s) only
and do not necessarily reflect those of the European Union. Neither
the European Union nor the granting authority can be held responsible for them. \\ 

\appendix 
% \bbbtext{\textit{AS: as given by PRX Quantum:
% Do not use Supplemental Material to avoid a length limit; often, a short paper accompanied by a lengthy supplement is not appropriate. Editors use their judgment to decide if a longer manuscript with all material integrated into the main text is required. The editors may seek guidance in this decision from the referees who review the manuscript and Supplemental Material. In a longer format manuscript, it may be best to present additional material as an appendix to the main article, rather than as Supplemental Material.}}

\section{\texorpdfstring{$(\isym{}$,$\qsym{})$}{(I,Q)}-space trajectory}
\label{app:varphi}

\begin{figure}
    \centering
\includegraphics[width=\columnwidth]{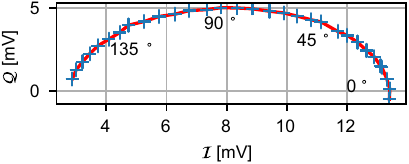}
    \caption{\textbf{$(\isym{}$,$\qsym{})$-space trajectory} Blue crosses indicate $\isym{}/\qsym{}$ values when a small $\brf$ of \SI{4}{\nano\tesla} is applied,
    %0.1 V peak-peak 8 nT peak-peak
    for a range of phase offsets $\varphi$ between the \rfsym{} field carrier and the pump modulation $R_\mathrm{OP}(t)$. Red curve is a linear interpolation between blue crosses.}
    \label{fig:iqspace}
\end{figure}

Fig.~\ref{fig:iqspace} shows an experimentally-measured $(\isym{}$,$\qsym{})$-space trajectory parameterized by $\varphi$, for $\delta\bdc{}=0$ and small $B_\mathrm{rf}$.
For  $\varphi=n\pi, n\in \mathbbm{Z}$, we have $\mathcal{Y}=0$, hence $\mathcal{X}$ encodes the amplitude of the \rf{} field, which is our operating parameter of interest. In the article we continue the description using amplitude of $B_\mathrm{rf}$ instead of $\mathcal{X}$.

\section{Threshold selection for receiver operation}
The threshold for the high/low logical level depends on the frequency channel. By observing a statistic of many transmissions, the channel-specific threshold is calibrated by taking the running average $V_H^{(j)}$ ($V_L^{(j)}$) with a window of $N_w=5$ logical symbols of the maxima (minima) within each symbol
\begin{equation}
    V_H^{(j)}=\max_{k\in  [j-\lfloor N_w/2\rfloor,j+\lfloor N_w/2 \rfloor]}  \langle\langle\isym{}(t)\rangle_{t_k}\rangle_{N_\mathrm{TX}},
\end{equation}
and by choosing the local threshold for the $j$-th symbol as a midpoint $V_\mathrm{thr}^{(j)}=(V_H^{(j)}+V_L^{(j)})/2$, where $\langle . \rangle_{t_k}$ denotes the average over the $k$-th symbol duration and $\langle . \rangle_{N_\mathrm{TX}}$ over the $N_\mathrm{TX}$ repetitions.

\section{External noise for receiver operation}
The feasibility of operation in a noisy environment (e.g. unshielded) is exemplified by adding Gaussian white noise to the $z$ component of the magnetic field $B_z$, independent of the feedback loop which alters the $B_x$ component. A measurement with white noise is performed only for $\brf{}=\SI{6.39}{\nano\tesla}$ and a noise level corresponding to a signal-to-noise ratio $\mathrm{SNR}=\brf{}/B_\mathrm{noise}$ of about 5, where $(B_\mathrm{noise})^2$ is the noise power spectral density integrated over the $\SI{100}{\hertz}$ transmission bandwidth. SNR was confirmed with OPM power spectral density measurements for a range of introduced noise powers.

%max 50 references
%\bibliographystyle{apsrev4-1no-url.bst}
\bibliography{main}% common bib file

\begin{thebibliography}{67}%
\makeatletter
\providecommand \@ifxundefined [1]{%
 \@ifx{#1\undefined}
}%
\providecommand \@ifnum [1]{%
 \ifnum #1\expandafter \@firstoftwo
 \else \expandafter \@secondoftwo
 \fi
}%
\providecommand \@ifx [1]{%
 \ifx #1\expandafter \@firstoftwo
 \else \expandafter \@secondoftwo
 \fi
}%
\providecommand \natexlab [1]{#1}%
\providecommand \enquote  [1]{``#1''}%
\providecommand \bibnamefont  [1]{#1}%
\providecommand \bibfnamefont [1]{#1}%
\providecommand \citenamefont [1]{#1}%
\providecommand \href@noop [0]{\@secondoftwo}%
\providecommand \href [0]{\begingroup \@sanitize@url \@href}%
\providecommand \@href[1]{\@@startlink{#1}\@@href}%
\providecommand \@@href[1]{\endgroup#1\@@endlink}%
\providecommand \@sanitize@url [0]{\catcode `\\12\catcode `\$12\catcode
  `\&12\catcode `\#12\catcode `\^12\catcode `\_12\catcode `\%12\relax}%
\providecommand \@@startlink[1]{}%
\providecommand \@@endlink[0]{}%
\providecommand \url  [0]{\begingroup\@sanitize@url \@url }%
\providecommand \@url [1]{\endgroup\@href {#1}{\urlprefix }}%
\providecommand \urlprefix  [0]{URL }%
\providecommand \Eprint [0]{\href }%
\providecommand \doibase [0]{https://doi.org/}%
\providecommand \selectlanguage [0]{\@gobble}%
\providecommand \bibinfo  [0]{\@secondoftwo}%
\providecommand \bibfield  [0]{\@secondoftwo}%
\providecommand \translation [1]{[#1]}%
\providecommand \BibitemOpen [0]{}%
\providecommand \bibitemStop [0]{}%
\providecommand \bibitemNoStop [0]{.\EOS\space}%
\providecommand \EOS [0]{\spacefactor3000\relax}%
\providecommand \BibitemShut  [1]{\csname bibitem#1\endcsname}%
\let\auto@bib@innerbib\@empty
%</preamble>
\bibitem [{\citenamefont {Helstrom}(1969)}]{Helstrom1969}%
  \BibitemOpen
  \bibfield  {author} {\bibinfo {author} {\bibfnamefont {C.~W.}\ \bibnamefont
  {Helstrom}},\ }\bibfield  {title} {\bibinfo {title} {Quantum detection and
  estimation theory},\ }\href {https://doi.org/10.1007/BF01007479} {\bibfield
  {journal} {\bibinfo  {journal} {J. Stat. Phys.}\ }\textbf {\bibinfo {volume}
  {1}},\ \bibinfo {pages} {231} (\bibinfo {year} {1969})}\BibitemShut {NoStop}%
\bibitem [{\citenamefont {Degen}\ \emph {et~al.}(2017)\citenamefont {Degen},
  \citenamefont {Reinhard},\ and\ \citenamefont {Cappellaro}}]{Degen2017}%
  \BibitemOpen
  \bibfield  {author} {\bibinfo {author} {\bibfnamefont {C.~L.}\ \bibnamefont
  {Degen}}, \bibinfo {author} {\bibfnamefont {F.}~\bibnamefont {Reinhard}},\
  and\ \bibinfo {author} {\bibfnamefont {P.}~\bibnamefont {Cappellaro}},\
  }\bibfield  {title} {\bibinfo {title} {Quantum sensing},\ }\href
  {https://doi.org/10.1103/RevModPhys.89.035002} {\bibfield  {journal}
  {\bibinfo  {journal} {Rev. Mod. Phys.}\ }\textbf {\bibinfo {volume} {89}},\
  \bibinfo {pages} {035002} (\bibinfo {year} {2017})}\BibitemShut {NoStop}%
\bibitem [{\citenamefont {Pezz\`e}\ \emph {et~al.}(2018)\citenamefont
  {Pezz\`e}, \citenamefont {Smerzi}, \citenamefont {Oberthaler}, \citenamefont
  {Schmied},\ and\ \citenamefont {Treutlein}}]{PezzeRMP2018}%
  \BibitemOpen
  \bibfield  {author} {\bibinfo {author} {\bibfnamefont {L.}~\bibnamefont
  {Pezz\`e}}, \bibinfo {author} {\bibfnamefont {A.}~\bibnamefont {Smerzi}},
  \bibinfo {author} {\bibfnamefont {M.~K.}\ \bibnamefont {Oberthaler}},
  \bibinfo {author} {\bibfnamefont {R.}~\bibnamefont {Schmied}},\ and\ \bibinfo
  {author} {\bibfnamefont {P.}~\bibnamefont {Treutlein}},\ }\bibfield  {title}
  {\bibinfo {title} {Quantum metrology with nonclassical states of atomic
  ensembles},\ }\href {https://doi.org/10.1103/RevModPhys.90.035005} {\bibfield
   {journal} {\bibinfo  {journal} {Rev. Mod. Phys.}\ }\textbf {\bibinfo
  {volume} {90}},\ \bibinfo {pages} {035005} (\bibinfo {year}
  {2018})}\BibitemShut {NoStop}%
\bibitem [{\citenamefont {Braun}\ \emph {et~al.}(2018)\citenamefont {Braun},
  \citenamefont {Adesso}, \citenamefont {Benatti}, \citenamefont {Floreanini},
  \citenamefont {Marzolino}, \citenamefont {Mitchell},\ and\ \citenamefont
  {Pirandola}}]{BraunRMP2018}%
  \BibitemOpen
  \bibfield  {author} {\bibinfo {author} {\bibfnamefont {D.}~\bibnamefont
  {Braun}}, \bibinfo {author} {\bibfnamefont {G.}~\bibnamefont {Adesso}},
  \bibinfo {author} {\bibfnamefont {F.}~\bibnamefont {Benatti}}, \bibinfo
  {author} {\bibfnamefont {R.}~\bibnamefont {Floreanini}}, \bibinfo {author}
  {\bibfnamefont {U.}~\bibnamefont {Marzolino}}, \bibinfo {author}
  {\bibfnamefont {M.~W.}\ \bibnamefont {Mitchell}},\ and\ \bibinfo {author}
  {\bibfnamefont {S.}~\bibnamefont {Pirandola}},\ }\bibfield  {title} {\bibinfo
  {title} {Quantum-enhanced measurements without entanglement},\ }\href
  {https://doi.org/10.1103/RevModPhys.90.035006} {\bibfield  {journal}
  {\bibinfo  {journal} {Rev. Mod. Phys.}\ }\textbf {\bibinfo {volume} {90}},\
  \bibinfo {pages} {035006} (\bibinfo {year} {2018})}\BibitemShut {NoStop}%
\bibitem [{\citenamefont {Caves}(1981{\natexlab{a}})}]{CavesPRD1981}%
  \BibitemOpen
  \bibfield  {author} {\bibinfo {author} {\bibfnamefont {C.~M.}\ \bibnamefont
  {Caves}},\ }\bibfield  {title} {\bibinfo {title} {Quantum-mechanical noise in
  an interferometer},\ }\href {https://doi.org/10.1103/PhysRevD.23.1693}
  {\bibfield  {journal} {\bibinfo  {journal} {Phys. Rev. D}\ }\textbf {\bibinfo
  {volume} {23}},\ \bibinfo {pages} {1693} (\bibinfo {year}
  {1981}{\natexlab{a}})}\BibitemShut {NoStop}%
\bibitem [{\citenamefont {Huelga}\ \emph {et~al.}(1997)\citenamefont {Huelga},
  \citenamefont {Macchiavello}, \citenamefont {Pellizzari}, \citenamefont
  {Ekert}, \citenamefont {Plenio},\ and\ \citenamefont
  {Cirac}}]{HuelgaPRL1997}%
  \BibitemOpen
  \bibfield  {author} {\bibinfo {author} {\bibfnamefont {S.~F.}\ \bibnamefont
  {Huelga}}, \bibinfo {author} {\bibfnamefont {C.}~\bibnamefont
  {Macchiavello}}, \bibinfo {author} {\bibfnamefont {T.}~\bibnamefont
  {Pellizzari}}, \bibinfo {author} {\bibfnamefont {A.~K.}\ \bibnamefont
  {Ekert}}, \bibinfo {author} {\bibfnamefont {M.~B.}\ \bibnamefont {Plenio}},\
  and\ \bibinfo {author} {\bibfnamefont {J.~I.}\ \bibnamefont {Cirac}},\
  }\bibfield  {title} {\bibinfo {title} {Improvement of frequency standards
  with quantum entanglement},\ }\href
  {https://doi.org/10.1103/PhysRevLett.79.3865} {\bibfield  {journal} {\bibinfo
   {journal} {Phys. Rev. Lett.}\ }\textbf {\bibinfo {volume} {79}},\ \bibinfo
  {pages} {3865} (\bibinfo {year} {1997})}\BibitemShut {NoStop}%
\bibitem [{\citenamefont {Orenes}\ \emph {et~al.}(2022)\citenamefont {Orenes},
  \citenamefont {Sewell}, \citenamefont {Lodewyck},\ and\ \citenamefont
  {Mitchell}}]{BenedictoPRL2022}%
  \BibitemOpen
  \bibfield  {author} {\bibinfo {author} {\bibfnamefont {D.~B.}\ \bibnamefont
  {Orenes}}, \bibinfo {author} {\bibfnamefont {R.~J.}\ \bibnamefont {Sewell}},
  \bibinfo {author} {\bibfnamefont {J.}~\bibnamefont {Lodewyck}},\ and\
  \bibinfo {author} {\bibfnamefont {M.~W.}\ \bibnamefont {Mitchell}},\
  }\bibfield  {title} {\bibinfo {title} {Improving short-term stability in
  optical lattice clocks by quantum nondemolition measurement},\ }\href
  {https://doi.org/10.1103/PhysRevLett.128.153201} {\bibfield  {journal}
  {\bibinfo  {journal} {Phys. Rev. Lett.}\ }\textbf {\bibinfo {volume} {128}},\
  \bibinfo {pages} {153201} (\bibinfo {year} {2022})}\BibitemShut {NoStop}%
\bibitem [{\citenamefont {Grangier}\ \emph {et~al.}(1987)\citenamefont
  {Grangier}, \citenamefont {Slusher}, \citenamefont {Yurke},\ and\
  \citenamefont {LaPorta}}]{GrangierPRL1987}%
  \BibitemOpen
  \bibfield  {author} {\bibinfo {author} {\bibfnamefont {P.}~\bibnamefont
  {Grangier}}, \bibinfo {author} {\bibfnamefont {R.~E.}\ \bibnamefont
  {Slusher}}, \bibinfo {author} {\bibfnamefont {B.}~\bibnamefont {Yurke}},\
  and\ \bibinfo {author} {\bibfnamefont {A.}~\bibnamefont {LaPorta}},\
  }\bibfield  {title} {\bibinfo {title} {Squeezed-light--enhanced polarization
  interferometer},\ }\href {https://doi.org/10.1103/PhysRevLett.59.2153}
  {\bibfield  {journal} {\bibinfo  {journal} {Phys. Rev. Lett.}\ }\textbf
  {\bibinfo {volume} {59}},\ \bibinfo {pages} {2153} (\bibinfo {year}
  {1987})}\BibitemShut {NoStop}%
\bibitem [{\citenamefont {Leroux}\ \emph {et~al.}(2010)\citenamefont {Leroux},
  \citenamefont {Schleier-Smith},\ and\ \citenamefont
  {Vuleti{\'c}}}]{LerouxPRL2010}%
  \BibitemOpen
  \bibfield  {author} {\bibinfo {author} {\bibfnamefont {I.~D.}\ \bibnamefont
  {Leroux}}, \bibinfo {author} {\bibfnamefont {M.~H.}\ \bibnamefont
  {Schleier-Smith}},\ and\ \bibinfo {author} {\bibfnamefont {V.}~\bibnamefont
  {Vuleti{\'c}}},\ }\bibfield  {title} {\bibinfo {title} {{Implementation of
  cavity squeezing of a collective atomic spin}},\ }\href@noop {} {\bibfield
  {journal} {\bibinfo  {journal} {Phys. Rev. Lett.}\ }\textbf {\bibinfo
  {volume} {104}},\ \bibinfo {pages} {73602} (\bibinfo {year}
  {2010})}\BibitemShut {NoStop}%
\bibitem [{\citenamefont {Hosten}\ \emph {et~al.}(2016)\citenamefont {Hosten},
  \citenamefont {Engelsen}, \citenamefont {Krishnakumar},\ and\ \citenamefont
  {Kasevich}}]{HostenN2016}%
  \BibitemOpen
  \bibfield  {author} {\bibinfo {author} {\bibfnamefont {O.}~\bibnamefont
  {Hosten}}, \bibinfo {author} {\bibfnamefont {N.~J.}\ \bibnamefont
  {Engelsen}}, \bibinfo {author} {\bibfnamefont {R.}~\bibnamefont
  {Krishnakumar}},\ and\ \bibinfo {author} {\bibfnamefont {M.~A.}\ \bibnamefont
  {Kasevich}},\ }\bibfield  {title} {\bibinfo {title} {Measurement noise 100
  times lower than the quantum-projection limit using entangled atoms},\ }\href
  {https://doi.org/10.1038/nature16176} {\bibfield  {journal} {\bibinfo
  {journal} {Nature}\ }\textbf {\bibinfo {volume} {529}},\ \bibinfo {pages}
  {505} (\bibinfo {year} {2016})}\BibitemShut {NoStop}%
\bibitem [{\citenamefont {Huang}(2019)}]{HuangThesis2019}%
  \BibitemOpen
  \bibfield  {author} {\bibinfo {author} {\bibfnamefont {M.-Z.}\ \bibnamefont
  {Huang}},\ }\emph {\bibinfo {title} {{Spin squeezing and spin dynamics in a
  trapped-atom clock}}},\ \href {https://tel.archives-ouvertes.fr/tel-02356949}
  {\bibinfo {type} {Theses}},\ \bibinfo  {school} {{Sorbonne Universit{\'e}}}
  (\bibinfo {year} {2019})\BibitemShut {NoStop}%
\bibitem [{\citenamefont {Wolfgramm}\ \emph {et~al.}(2010)\citenamefont
  {Wolfgramm}, \citenamefont {Cer\`e}, \citenamefont {Beduini}, \citenamefont
  {Predojevi\ifmmode~\acute{c}\else \'{c}\fi{}}, \citenamefont {Koschorreck},\
  and\ \citenamefont {Mitchell}}]{WolfgrammPRL2010}%
  \BibitemOpen
  \bibfield  {author} {\bibinfo {author} {\bibfnamefont {F.}~\bibnamefont
  {Wolfgramm}}, \bibinfo {author} {\bibfnamefont {A.}~\bibnamefont {Cer\`e}},
  \bibinfo {author} {\bibfnamefont {F.~A.}\ \bibnamefont {Beduini}}, \bibinfo
  {author} {\bibfnamefont {A.}~\bibnamefont {Predojevi\ifmmode~\acute{c}\else
  \'{c}\fi{}}}, \bibinfo {author} {\bibfnamefont {M.}~\bibnamefont
  {Koschorreck}},\ and\ \bibinfo {author} {\bibfnamefont {M.~W.}\ \bibnamefont
  {Mitchell}},\ }\bibfield  {title} {\bibinfo {title} {Squeezed-light optical
  magnetometry},\ }\href {https://doi.org/10.1103/PhysRevLett.105.053601}
  {\bibfield  {journal} {\bibinfo  {journal} {Phys. Rev. Lett.}\ }\textbf
  {\bibinfo {volume} {105}},\ \bibinfo {pages} {053601} (\bibinfo {year}
  {2010})}\BibitemShut {NoStop}%
\bibitem [{\citenamefont {Wasilewski}\ \emph {et~al.}(2010)\citenamefont
  {Wasilewski}, \citenamefont {Jensen}, \citenamefont {Krauter}, \citenamefont
  {Renema}, \citenamefont {Balabas},\ and\ \citenamefont
  {Polzik}}]{Wasilewski2010}%
  \BibitemOpen
  \bibfield  {author} {\bibinfo {author} {\bibfnamefont {W.}~\bibnamefont
  {Wasilewski}}, \bibinfo {author} {\bibfnamefont {K.}~\bibnamefont {Jensen}},
  \bibinfo {author} {\bibfnamefont {H.}~\bibnamefont {Krauter}}, \bibinfo
  {author} {\bibfnamefont {J.~J.}\ \bibnamefont {Renema}}, \bibinfo {author}
  {\bibfnamefont {M.~V.}\ \bibnamefont {Balabas}},\ and\ \bibinfo {author}
  {\bibfnamefont {E.~S.}\ \bibnamefont {Polzik}},\ }\bibfield  {title}
  {\bibinfo {title} {Quantum noise limited and entanglement-assisted
  magnetometry},\ }\href {https://doi.org/10.1103/PhysRevLett.104.133601}
  {\bibfield  {journal} {\bibinfo  {journal} {Phys. Rev. Lett.}\ }\textbf
  {\bibinfo {volume} {104}},\ \bibinfo {pages} {133601} (\bibinfo {year}
  {2010})}\BibitemShut {NoStop}%
\bibitem [{\citenamefont {Horrom}\ \emph {et~al.}(2012)\citenamefont {Horrom},
  \citenamefont {Singh}, \citenamefont {Dowling},\ and\ \citenamefont
  {Mikhailov}}]{HorromPRA2012}%
  \BibitemOpen
  \bibfield  {author} {\bibinfo {author} {\bibfnamefont {T.}~\bibnamefont
  {Horrom}}, \bibinfo {author} {\bibfnamefont {R.}~\bibnamefont {Singh}},
  \bibinfo {author} {\bibfnamefont {J.~P.}\ \bibnamefont {Dowling}},\ and\
  \bibinfo {author} {\bibfnamefont {E.~E.}\ \bibnamefont {Mikhailov}},\
  }\bibfield  {title} {\bibinfo {title} {Quantum-enhanced magnetometer with
  low-frequency squeezing},\ }\href
  {https://doi.org/10.1103/PhysRevA.86.023803} {\bibfield  {journal} {\bibinfo
  {journal} {Phys. Rev. A}\ }\textbf {\bibinfo {volume} {86}},\ \bibinfo
  {pages} {023803} (\bibinfo {year} {2012})}\BibitemShut {NoStop}%
\bibitem [{\citenamefont {Sewell}\ \emph {et~al.}(2012)\citenamefont {Sewell},
  \citenamefont {Koschorreck}, \citenamefont {Napolitano}, \citenamefont
  {Dubost}, \citenamefont {Behbood},\ and\ \citenamefont
  {Mitchell}}]{Sewell2012}%
  \BibitemOpen
  \bibfield  {author} {\bibinfo {author} {\bibfnamefont {R.~J.}\ \bibnamefont
  {Sewell}}, \bibinfo {author} {\bibfnamefont {M.}~\bibnamefont {Koschorreck}},
  \bibinfo {author} {\bibfnamefont {M.}~\bibnamefont {Napolitano}}, \bibinfo
  {author} {\bibfnamefont {B.}~\bibnamefont {Dubost}}, \bibinfo {author}
  {\bibfnamefont {N.}~\bibnamefont {Behbood}},\ and\ \bibinfo {author}
  {\bibfnamefont {M.~W.}\ \bibnamefont {Mitchell}},\ }\bibfield  {title}
  {\bibinfo {title} {Magnetic sensitivity beyond the projection noise limit by
  spin squeezing},\ }\href {https://doi.org/10.1103/PhysRevLett.109.253605}
  {\bibfield  {journal} {\bibinfo  {journal} {Phys. Rev. Lett.}\ }\textbf
  {\bibinfo {volume} {109}},\ \bibinfo {pages} {253605} (\bibinfo {year}
  {2012})}\BibitemShut {NoStop}%
\bibitem [{\citenamefont {Aasi}\ \emph {et~al.}(2013)\citenamefont {Aasi},
  \citenamefont {Abadie}, \citenamefont {Abbott}, \citenamefont {Abbott},
  \citenamefont {Abbott}, \citenamefont {Abernathy}, \citenamefont {Adams},
  \citenamefont {Adams}, \citenamefont {Addesso}, \citenamefont {Adhikari}
  \emph {et~al.}}]{Aasi2013NP}%
  \BibitemOpen
  \bibfield  {author} {\bibinfo {author} {\bibfnamefont {J.}~\bibnamefont
  {Aasi}}, \bibinfo {author} {\bibfnamefont {J.}~\bibnamefont {Abadie}},
  \bibinfo {author} {\bibfnamefont {B.}~\bibnamefont {Abbott}}, \bibinfo
  {author} {\bibfnamefont {R.}~\bibnamefont {Abbott}}, \bibinfo {author}
  {\bibfnamefont {T.}~\bibnamefont {Abbott}}, \bibinfo {author} {\bibfnamefont
  {M.}~\bibnamefont {Abernathy}}, \bibinfo {author} {\bibfnamefont
  {C.}~\bibnamefont {Adams}}, \bibinfo {author} {\bibfnamefont
  {T.}~\bibnamefont {Adams}}, \bibinfo {author} {\bibfnamefont
  {P.}~\bibnamefont {Addesso}}, \bibinfo {author} {\bibfnamefont
  {R.}~\bibnamefont {Adhikari}}, \emph {et~al.},\ }\bibfield  {title} {\bibinfo
  {title} {Enhanced sensitivity of the {LIGO} gravitational wave detector by
  using squeezed states of light},\ }\href@noop {} {\bibfield  {journal}
  {\bibinfo  {journal} {Nat. Photonics}\ }\textbf {\bibinfo {volume} {7}},\
  \bibinfo {pages} {613} (\bibinfo {year} {2013})}\BibitemShut {NoStop}%
\bibitem [{\citenamefont {Tse~{\it et al.}}(2019)}]{TsePRL2019Short}%
  \BibitemOpen
  \bibfield  {author} {\bibinfo {author} {\bibfnamefont {M.}~\bibnamefont
  {Tse~{\it et al.}}},\ }\bibfield  {title} {\bibinfo {title} {Quantum-enhanced
  advanced {LIGO} detectors in the era of gravitational-wave astronomy},\
  }\href {https://doi.org/10.1103/PhysRevLett.123.231107} {\bibfield  {journal}
  {\bibinfo  {journal} {Phys. Rev. Lett.}\ }\textbf {\bibinfo {volume} {123}},\
  \bibinfo {pages} {231107} (\bibinfo {year} {2019})}\BibitemShut {NoStop}%
\bibitem [{\citenamefont {Acernese}\ and\ \citenamefont {et.
  al.}(2020)}]{AcernesePRL2020}%
  \BibitemOpen
  \bibfield  {author} {\bibinfo {author} {\bibfnamefont {F.}~\bibnamefont
  {Acernese}}\ and\ \bibinfo {author} {\bibnamefont {et. al.}} (\bibinfo
  {collaboration} {The Virgo Collaboration}),\ }\bibfield  {title} {\bibinfo
  {title} {Quantum backaction on kg-scale mirrors: Observation of radiation
  pressure noise in the advanced virgo detector},\ }\href
  {https://doi.org/10.1103/PhysRevLett.125.131101} {\bibfield  {journal}
  {\bibinfo  {journal} {Phys. Rev. Lett.}\ }\textbf {\bibinfo {volume} {125}},\
  \bibinfo {pages} {131101} (\bibinfo {year} {2020})}\BibitemShut {NoStop}%
\bibitem [{\citenamefont {Troullinou}\ \emph
  {et~al.}(2021{\natexlab{a}})\citenamefont {Troullinou}, \citenamefont
  {Jim\'enez-Mart\'{\i}nez}, \citenamefont {Kong}, \citenamefont {Lucivero},\
  and\ \citenamefont {Mitchell}}]{Troullinou2021}%
  \BibitemOpen
  \bibfield  {author} {\bibinfo {author} {\bibfnamefont {C.}~\bibnamefont
  {Troullinou}}, \bibinfo {author} {\bibfnamefont {R.}~\bibnamefont
  {Jim\'enez-Mart\'{\i}nez}}, \bibinfo {author} {\bibfnamefont
  {J.}~\bibnamefont {Kong}}, \bibinfo {author} {\bibfnamefont {V.~G.}\
  \bibnamefont {Lucivero}},\ and\ \bibinfo {author} {\bibfnamefont {M.~W.}\
  \bibnamefont {Mitchell}},\ }\bibfield  {title} {\bibinfo {title}
  {Squeezed-light enhancement and backaction evasion in a high sensitivity
  optically pumped magnetometer},\ }\href
  {https://doi.org/10.1103/PhysRevLett.127.193601} {\bibfield  {journal}
  {\bibinfo  {journal} {Phys. Rev. Lett.}\ }\textbf {\bibinfo {volume} {127}},\
  \bibinfo {pages} {193601} (\bibinfo {year} {2021}{\natexlab{a}})}\BibitemShut
  {NoStop}%
\bibitem [{\citenamefont {Zheng}\ \emph {et~al.}(2023)\citenamefont {Zheng},
  \citenamefont {Wang}, \citenamefont {Schmieg}, \citenamefont {Oesterle},\
  and\ \citenamefont {Polzik}}]{ZhengPRL2023}%
  \BibitemOpen
  \bibfield  {author} {\bibinfo {author} {\bibfnamefont {W.}~\bibnamefont
  {Zheng}}, \bibinfo {author} {\bibfnamefont {H.}~\bibnamefont {Wang}},
  \bibinfo {author} {\bibfnamefont {R.}~\bibnamefont {Schmieg}}, \bibinfo
  {author} {\bibfnamefont {A.}~\bibnamefont {Oesterle}},\ and\ \bibinfo
  {author} {\bibfnamefont {E.~S.}\ \bibnamefont {Polzik}},\ }\bibfield  {title}
  {\bibinfo {title} {Entanglement-enhanced magnetic induction tomography},\
  }\href {https://doi.org/10.1103/PhysRevLett.130.203602} {\bibfield  {journal}
  {\bibinfo  {journal} {Phys. Rev. Lett.}\ }\textbf {\bibinfo {volume} {130}},\
  \bibinfo {pages} {203602} (\bibinfo {year} {2023})}\BibitemShut {NoStop}%
\bibitem [{\citenamefont {McCuller}\ \emph {et~al.}(2020)\citenamefont
  {McCuller}, \citenamefont {Whittle}, \citenamefont {Ganapathy}, \citenamefont
  {Komori}, \citenamefont {Tse}, \citenamefont {Fernandez-Galiana},
  \citenamefont {Barsotti}, \citenamefont {Fritschel}, \citenamefont
  {MacInnis}, \citenamefont {Matichard}, \citenamefont {Mason}, \citenamefont
  {Mavalvala}, \citenamefont {Mittleman}, \citenamefont {Yu}, \citenamefont
  {Zucker},\ and\ \citenamefont {Evans}}]{McCullerPRL2020}%
  \BibitemOpen
  \bibfield  {author} {\bibinfo {author} {\bibfnamefont {L.}~\bibnamefont
  {McCuller}}, \bibinfo {author} {\bibfnamefont {C.}~\bibnamefont {Whittle}},
  \bibinfo {author} {\bibfnamefont {D.}~\bibnamefont {Ganapathy}}, \bibinfo
  {author} {\bibfnamefont {K.}~\bibnamefont {Komori}}, \bibinfo {author}
  {\bibfnamefont {M.}~\bibnamefont {Tse}}, \bibinfo {author} {\bibfnamefont
  {A.}~\bibnamefont {Fernandez-Galiana}}, \bibinfo {author} {\bibfnamefont
  {L.}~\bibnamefont {Barsotti}}, \bibinfo {author} {\bibfnamefont
  {P.}~\bibnamefont {Fritschel}}, \bibinfo {author} {\bibfnamefont
  {M.}~\bibnamefont {MacInnis}}, \bibinfo {author} {\bibfnamefont
  {F.}~\bibnamefont {Matichard}}, \bibinfo {author} {\bibfnamefont
  {K.}~\bibnamefont {Mason}}, \bibinfo {author} {\bibfnamefont
  {N.}~\bibnamefont {Mavalvala}}, \bibinfo {author} {\bibfnamefont
  {R.}~\bibnamefont {Mittleman}}, \bibinfo {author} {\bibfnamefont
  {H.}~\bibnamefont {Yu}}, \bibinfo {author} {\bibfnamefont {M.~E.}\
  \bibnamefont {Zucker}},\ and\ \bibinfo {author} {\bibfnamefont
  {M.}~\bibnamefont {Evans}},\ }\bibfield  {title} {\bibinfo {title}
  {Frequency-dependent squeezing for advanced {LIGO}},\ }\href
  {https://doi.org/10.1103/PhysRevLett.124.171102} {\bibfield  {journal}
  {\bibinfo  {journal} {Phys. Rev. Lett.}\ }\textbf {\bibinfo {volume} {124}},\
  \bibinfo {pages} {171102} (\bibinfo {year} {2020})}\BibitemShut {NoStop}%
\bibitem [{\citenamefont {McCuller}\ and\ \citenamefont
  {et.al.}(2021)}]{McCuller2021PRD}%
  \BibitemOpen
  \bibfield  {author} {\bibinfo {author} {\bibfnamefont {L.}~\bibnamefont
  {McCuller}}\ and\ \bibinfo {author} {\bibnamefont {et.al.}},\ }\bibfield
  {title} {\bibinfo {title} {{LIGO}'s quantum response to squeezed states},\
  }\href {https://doi.org/10.1103/PhysRevD.104.062006} {\bibfield  {journal}
  {\bibinfo  {journal} {Phys. Rev. D}\ }\textbf {\bibinfo {volume} {104}},\
  \bibinfo {pages} {062006} (\bibinfo {year} {2021})}\BibitemShut {NoStop}%
\bibitem [{\citenamefont {Szczykulska}\ \emph {et~al.}(2016)\citenamefont
  {Szczykulska}, \citenamefont {Baumgratz},\ and\ \citenamefont
  {Datta}}]{SzczykulskaAPX2016}%
  \BibitemOpen
  \bibfield  {author} {\bibinfo {author} {\bibfnamefont {M.}~\bibnamefont
  {Szczykulska}}, \bibinfo {author} {\bibfnamefont {T.}~\bibnamefont
  {Baumgratz}},\ and\ \bibinfo {author} {\bibfnamefont {A.}~\bibnamefont
  {Datta}},\ }\bibfield  {title} {\bibinfo {title} {Multi-parameter quantum
  metrology},\ }\href {https://doi.org/10.1080/23746149.2016.1230476}
  {\bibfield  {journal} {\bibinfo  {journal} {Adv. Phys.: X}\ }\textbf
  {\bibinfo {volume} {1}},\ \bibinfo {pages} {621} (\bibinfo {year}
  {2016})}\BibitemShut {NoStop}%
\bibitem [{\citenamefont {Demkowicz-Dobrza{\'{n}}ski}\ \emph
  {et~al.}(2020)\citenamefont {Demkowicz-Dobrza{\'{n}}ski}, \citenamefont
  {G{\'{o}}recki},\ and\ \citenamefont {Gu{\c{t}}{\u{a}}}}]{Demkowicz2020}%
  \BibitemOpen
  \bibfield  {author} {\bibinfo {author} {\bibfnamefont {R.}~\bibnamefont
  {Demkowicz-Dobrza{\'{n}}ski}}, \bibinfo {author} {\bibfnamefont
  {W.}~\bibnamefont {G{\'{o}}recki}},\ and\ \bibinfo {author} {\bibfnamefont
  {M.}~\bibnamefont {Gu{\c{t}}{\u{a}}}},\ }\bibfield  {title} {\bibinfo {title}
  {Multi-parameter estimation beyond quantum fisher information},\ }\href
  {https://doi.org/10.1088/1751-8121/ab8ef3} {\bibfield  {journal} {\bibinfo
  {journal} {J. Phys. A Math. Theor.}\ }\textbf {\bibinfo {volume} {53}},\
  \bibinfo {pages} {363001} (\bibinfo {year} {2020})}\BibitemShut {NoStop}%
\bibitem [{\citenamefont {Proctor}\ \emph {et~al.}(2018)\citenamefont
  {Proctor}, \citenamefont {Knott},\ and\ \citenamefont
  {Dunningham}}]{ProctorPRL2018}%
  \BibitemOpen
  \bibfield  {author} {\bibinfo {author} {\bibfnamefont {T.~J.}\ \bibnamefont
  {Proctor}}, \bibinfo {author} {\bibfnamefont {P.~A.}\ \bibnamefont {Knott}},\
  and\ \bibinfo {author} {\bibfnamefont {J.~A.}\ \bibnamefont {Dunningham}},\
  }\bibfield  {title} {\bibinfo {title} {Multiparameter estimation in networked
  quantum sensors},\ }\href {https://doi.org/10.1103/PhysRevLett.120.080501}
  {\bibfield  {journal} {\bibinfo  {journal} {Phys. Rev. Lett.}\ }\textbf
  {\bibinfo {volume} {120}},\ \bibinfo {pages} {080501} (\bibinfo {year}
  {2018})}\BibitemShut {NoStop}%
\bibitem [{\citenamefont {{\v R}eha{\v c}ek}\ \emph {et~al.}(2017)\citenamefont
  {{\v R}eha{\v c}ek}, \citenamefont {Hradil}, \citenamefont {Stoklasa},
  \citenamefont {Pa{\'u}r}, \citenamefont {Grover}, \citenamefont {Krzic},\
  and\ \citenamefont {S{\'a}nchez-Soto}}]{RehacekPRA2017}%
  \BibitemOpen
  \bibfield  {author} {\bibinfo {author} {\bibfnamefont {J.}~\bibnamefont {{\v
  R}eha{\v c}ek}}, \bibinfo {author} {\bibfnamefont {Z.}~\bibnamefont
  {Hradil}}, \bibinfo {author} {\bibfnamefont {B.}~\bibnamefont {Stoklasa}},
  \bibinfo {author} {\bibfnamefont {M.}~\bibnamefont {Pa{\'u}r}}, \bibinfo
  {author} {\bibfnamefont {J.}~\bibnamefont {Grover}}, \bibinfo {author}
  {\bibfnamefont {A.}~\bibnamefont {Krzic}},\ and\ \bibinfo {author}
  {\bibfnamefont {L.~L.}\ \bibnamefont {S{\'a}nchez-Soto}},\ }\bibfield
  {title} {\bibinfo {title} {Multiparameter quantum metrology of incoherent
  point sources: Towards realistic superresolution},\ }\href
  {https://doi.org/10.1103/PhysRevA.96.062107} {\bibfield  {journal} {\bibinfo
  {journal} {Phys. Rev. A}\ }\textbf {\bibinfo {volume} {96}},\ \bibinfo
  {pages} {062107} (\bibinfo {year} {2017})}\BibitemShut {NoStop}%
\bibitem [{\citenamefont {Carollo}\ \emph {et~al.}(2019)\citenamefont
  {Carollo}, \citenamefont {Spagnolo}, \citenamefont {Dubkov},\ and\
  \citenamefont {Valenti}}]{CarolloJSMTE2019}%
  \BibitemOpen
  \bibfield  {author} {\bibinfo {author} {\bibfnamefont {A.}~\bibnamefont
  {Carollo}}, \bibinfo {author} {\bibfnamefont {B.}~\bibnamefont {Spagnolo}},
  \bibinfo {author} {\bibfnamefont {A.~A.}\ \bibnamefont {Dubkov}},\ and\
  \bibinfo {author} {\bibfnamefont {D.}~\bibnamefont {Valenti}},\ }\bibfield
  {title} {\bibinfo {title} {On quantumness in multi-parameter quantum
  estimation},\ }\href {https://doi.org/10.1088/1742-5468/ab3ccb} {\bibfield
  {journal} {\bibinfo  {journal} {J. Stat. Mech. Theory Exp.}\ }\textbf
  {\bibinfo {volume} {2019}},\ \bibinfo {pages} {094010} (\bibinfo {year}
  {2019})}\BibitemShut {NoStop}%
\bibitem [{\citenamefont {Liu}\ \emph {et~al.}(2019)\citenamefont {Liu},
  \citenamefont {Yuan}, \citenamefont {Lu},\ and\ \citenamefont
  {Wang}}]{LiuJPAMT2020}%
  \BibitemOpen
  \bibfield  {author} {\bibinfo {author} {\bibfnamefont {J.}~\bibnamefont
  {Liu}}, \bibinfo {author} {\bibfnamefont {H.}~\bibnamefont {Yuan}}, \bibinfo
  {author} {\bibfnamefont {X.-M.}\ \bibnamefont {Lu}},\ and\ \bibinfo {author}
  {\bibfnamefont {X.}~\bibnamefont {Wang}},\ }\bibfield  {title} {\bibinfo
  {title} {Quantum fisher information matrix and multiparameter estimation},\
  }\href {https://doi.org/10.1088/1751-8121/ab5d4d} {\bibfield  {journal}
  {\bibinfo  {journal} {J. Phys. A Math. Theor.}\ }\textbf {\bibinfo {volume}
  {53}},\ \bibinfo {pages} {023001} (\bibinfo {year} {2019})}\BibitemShut
  {NoStop}%
\bibitem [{\citenamefont {Huang}\ \emph {et~al.}(2021)\citenamefont {Huang},
  \citenamefont {Lupo},\ and\ \citenamefont {Kok}}]{HuangPRXQ2021}%
  \BibitemOpen
  \bibfield  {author} {\bibinfo {author} {\bibfnamefont {Z.}~\bibnamefont
  {Huang}}, \bibinfo {author} {\bibfnamefont {C.}~\bibnamefont {Lupo}},\ and\
  \bibinfo {author} {\bibfnamefont {P.}~\bibnamefont {Kok}},\ }\bibfield
  {title} {\bibinfo {title} {Quantum-limited estimation of range and
  velocity},\ }\href {https://doi.org/10.1103/PRXQuantum.2.030303} {\bibfield
  {journal} {\bibinfo  {journal} {PRX Quantum}\ }\textbf {\bibinfo {volume}
  {2}},\ \bibinfo {pages} {030303} (\bibinfo {year} {2021})}\BibitemShut
  {NoStop}%
\bibitem [{\citenamefont {Kaubruegger}\ \emph {et~al.}(2023)\citenamefont
  {Kaubruegger}, \citenamefont {Shankar}, \citenamefont {Vasilyev},\ and\
  \citenamefont {Zoller}}]{KaubrueggerPRXQ2023}%
  \BibitemOpen
  \bibfield  {author} {\bibinfo {author} {\bibfnamefont {R.}~\bibnamefont
  {Kaubruegger}}, \bibinfo {author} {\bibfnamefont {A.}~\bibnamefont
  {Shankar}}, \bibinfo {author} {\bibfnamefont {D.~V.}\ \bibnamefont
  {Vasilyev}},\ and\ \bibinfo {author} {\bibfnamefont {P.}~\bibnamefont
  {Zoller}},\ }\bibfield  {title} {\bibinfo {title} {Optimal and variational
  multiparameter quantum metrology and vector-field sensing},\ }\href
  {https://doi.org/10.1103/PRXQuantum.4.020333} {\bibfield  {journal} {\bibinfo
   {journal} {PRX Quantum}\ }\textbf {\bibinfo {volume} {4}},\ \bibinfo {pages}
  {020333} (\bibinfo {year} {2023})}\BibitemShut {NoStop}%
\bibitem [{\citenamefont {G\'orecki}\ and\ \citenamefont
  {Demkowicz-Dobrza\ifmmode~\acute{n}\else \'{n}\fi{}ski}(2022)}]{Gorecki2022}%
  \BibitemOpen
  \bibfield  {author} {\bibinfo {author} {\bibfnamefont {W.}~\bibnamefont
  {G\'orecki}}\ and\ \bibinfo {author} {\bibfnamefont {R.}~\bibnamefont
  {Demkowicz-Dobrza\ifmmode~\acute{n}\else \'{n}\fi{}ski}},\ }\bibfield
  {title} {\bibinfo {title} {Multiparameter quantum metrology in the heisenberg
  limit regime: Many-repetition scenario versus full optimization},\ }\href
  {https://doi.org/10.1103/PhysRevA.106.012424} {\bibfield  {journal} {\bibinfo
   {journal} {Phys. Rev. A}\ }\textbf {\bibinfo {volume} {106}},\ \bibinfo
  {pages} {012424} (\bibinfo {year} {2022})}\BibitemShut {NoStop}%
\bibitem [{\citenamefont {Colangelo}\ \emph
  {et~al.}(2017{\natexlab{a}})\citenamefont {Colangelo}, \citenamefont
  {Ciurana}, \citenamefont {Bianchet}, \citenamefont {Sewell},\ and\
  \citenamefont {Mitchell}}]{ColangeloN2017}%
  \BibitemOpen
  \bibfield  {author} {\bibinfo {author} {\bibfnamefont {G.}~\bibnamefont
  {Colangelo}}, \bibinfo {author} {\bibfnamefont {F.~M.}\ \bibnamefont
  {Ciurana}}, \bibinfo {author} {\bibfnamefont {L.~C.}\ \bibnamefont
  {Bianchet}}, \bibinfo {author} {\bibfnamefont {R.~J.}\ \bibnamefont
  {Sewell}},\ and\ \bibinfo {author} {\bibfnamefont {M.~W.}\ \bibnamefont
  {Mitchell}},\ }\bibfield  {title} {\bibinfo {title} {Simultaneous tracking of
  spin angle and amplitude beyond classical limits},\ }\href
  {https://doi.org/10.1038/nature21434} {\bibfield  {journal} {\bibinfo
  {journal} {Nature}\ }\textbf {\bibinfo {volume} {543}},\ \bibinfo {pages}
  {525} (\bibinfo {year} {2017}{\natexlab{a}})}\BibitemShut {NoStop}%
\bibitem [{\citenamefont {M{\o}ller}\ \emph {et~al.}(2017)\citenamefont
  {M{\o}ller}, \citenamefont {Thomas}, \citenamefont {Vasilakis}, \citenamefont
  {Zeuthen}, \citenamefont {Tsaturyan}, \citenamefont {Balabas}, \citenamefont
  {Jensen}, \citenamefont {Schliesser}, \citenamefont {Hammerer},\ and\
  \citenamefont {Polzik}}]{MollerN2017}%
  \BibitemOpen
  \bibfield  {author} {\bibinfo {author} {\bibfnamefont {C.~B.}\ \bibnamefont
  {M{\o}ller}}, \bibinfo {author} {\bibfnamefont {R.~A.}\ \bibnamefont
  {Thomas}}, \bibinfo {author} {\bibfnamefont {G.}~\bibnamefont {Vasilakis}},
  \bibinfo {author} {\bibfnamefont {E.}~\bibnamefont {Zeuthen}}, \bibinfo
  {author} {\bibfnamefont {Y.}~\bibnamefont {Tsaturyan}}, \bibinfo {author}
  {\bibfnamefont {M.}~\bibnamefont {Balabas}}, \bibinfo {author} {\bibfnamefont
  {K.}~\bibnamefont {Jensen}}, \bibinfo {author} {\bibfnamefont
  {A.}~\bibnamefont {Schliesser}}, \bibinfo {author} {\bibfnamefont
  {K.}~\bibnamefont {Hammerer}},\ and\ \bibinfo {author} {\bibfnamefont
  {E.~S.}\ \bibnamefont {Polzik}},\ }\bibfield  {title} {\bibinfo {title}
  {Quantum back-action-evading measurement of motion in a negative mass
  reference frame},\ }\href {https://doi.org/10.1038/nature22980} {\bibfield
  {journal} {\bibinfo  {journal} {Nature}\ }\textbf {\bibinfo {volume} {547}},\
  \bibinfo {pages} {191} (\bibinfo {year} {2017})}\BibitemShut {NoStop}%
\bibitem [{\citenamefont {Polino}\ \emph {et~al.}(2019)\citenamefont {Polino},
  \citenamefont {Riva}, \citenamefont {Valeri}, \citenamefont {Silvestri},
  \citenamefont {Corrielli}, \citenamefont {Crespi}, \citenamefont {Spagnolo},
  \citenamefont {Osellame},\ and\ \citenamefont {Sciarrino}}]{PolinoO2019}%
  \BibitemOpen
  \bibfield  {author} {\bibinfo {author} {\bibfnamefont {E.}~\bibnamefont
  {Polino}}, \bibinfo {author} {\bibfnamefont {M.}~\bibnamefont {Riva}},
  \bibinfo {author} {\bibfnamefont {M.}~\bibnamefont {Valeri}}, \bibinfo
  {author} {\bibfnamefont {R.}~\bibnamefont {Silvestri}}, \bibinfo {author}
  {\bibfnamefont {G.}~\bibnamefont {Corrielli}}, \bibinfo {author}
  {\bibfnamefont {A.}~\bibnamefont {Crespi}}, \bibinfo {author} {\bibfnamefont
  {N.}~\bibnamefont {Spagnolo}}, \bibinfo {author} {\bibfnamefont
  {R.}~\bibnamefont {Osellame}},\ and\ \bibinfo {author} {\bibfnamefont
  {F.}~\bibnamefont {Sciarrino}},\ }\bibfield  {title} {\bibinfo {title}
  {Experimental multiphase estimation on a chip},\ }\href
  {https://doi.org/10.1364/OPTICA.6.000288} {\bibfield  {journal} {\bibinfo
  {journal} {Optica}\ }\textbf {\bibinfo {volume} {6}},\ \bibinfo {pages} {288}
  (\bibinfo {year} {2019})}\BibitemShut {NoStop}%
\bibitem [{\citenamefont {Bai}\ \emph {et~al.}(2021)\citenamefont {Bai},
  \citenamefont {Wen}, \citenamefont {Yang}, \citenamefont {Zhang},
  \citenamefont {He}, \citenamefont {Wang},\ and\ \citenamefont
  {Wang}}]{bai2021quantum}%
  \BibitemOpen
  \bibfield  {author} {\bibinfo {author} {\bibfnamefont {L.}~\bibnamefont
  {Bai}}, \bibinfo {author} {\bibfnamefont {X.}~\bibnamefont {Wen}}, \bibinfo
  {author} {\bibfnamefont {Y.}~\bibnamefont {Yang}}, \bibinfo {author}
  {\bibfnamefont {L.}~\bibnamefont {Zhang}}, \bibinfo {author} {\bibfnamefont
  {J.}~\bibnamefont {He}}, \bibinfo {author} {\bibfnamefont {Y.}~\bibnamefont
  {Wang}},\ and\ \bibinfo {author} {\bibfnamefont {J.}~\bibnamefont {Wang}},\
  }\bibfield  {title} {\bibinfo {title} {Quantum-enhanced rubidium atomic
  magnetometer based on faraday rotation via 795 nm stokes operator squeezed
  light},\ }\href@noop {} {\bibfield  {journal} {\bibinfo  {journal} {Journal
  of Optics}\ }\textbf {\bibinfo {volume} {23}},\ \bibinfo {pages} {085202}
  (\bibinfo {year} {2021})}\BibitemShut {NoStop}%
\bibitem [{\citenamefont {Colangelo}\ \emph
  {et~al.}(2017{\natexlab{b}})\citenamefont {Colangelo}, \citenamefont
  {Ciurana}, \citenamefont {Bianchet}, \citenamefont {Sewell},\ and\
  \citenamefont {Mitchell}}]{Colangelo2017}%
  \BibitemOpen
  \bibfield  {author} {\bibinfo {author} {\bibfnamefont {G.}~\bibnamefont
  {Colangelo}}, \bibinfo {author} {\bibfnamefont {F.~M.}\ \bibnamefont
  {Ciurana}}, \bibinfo {author} {\bibfnamefont {L.~C.}\ \bibnamefont
  {Bianchet}}, \bibinfo {author} {\bibfnamefont {R.~J.}\ \bibnamefont
  {Sewell}},\ and\ \bibinfo {author} {\bibfnamefont {M.~W.}\ \bibnamefont
  {Mitchell}},\ }\bibfield  {title} {\bibinfo {title} {Simultaneous tracking of
  spin angle and amplitude beyond classical limits},\ }\href
  {https://doi.org/10.1038/nature21434} {\bibfield  {journal} {\bibinfo
  {journal} {Nature}\ }\textbf {\bibinfo {volume} {543}},\ \bibinfo {pages}
  {525} (\bibinfo {year} {2017}{\natexlab{b}})}\BibitemShut {NoStop}%
\bibitem [{\citenamefont {Mitchell}\ \emph {et~al.}(2004)\citenamefont
  {Mitchell}, \citenamefont {Lundeen},\ and\ \citenamefont
  {Steinberg}}]{MitchellN2004}%
  \BibitemOpen
  \bibfield  {author} {\bibinfo {author} {\bibfnamefont {M.~W.}\ \bibnamefont
  {Mitchell}}, \bibinfo {author} {\bibfnamefont {J.~S.}\ \bibnamefont
  {Lundeen}},\ and\ \bibinfo {author} {\bibfnamefont {A.~M.}\ \bibnamefont
  {Steinberg}},\ }\bibfield  {title} {\bibinfo {title} {Super-resolving phase
  measurements with a multiphoton entangled state},\ }\href
  {https://doi.org/10.1038/nature02493} {\bibfield  {journal} {\bibinfo
  {journal} {Nature}\ }\textbf {\bibinfo {volume} {429}},\ \bibinfo {pages}
  {161} (\bibinfo {year} {2004})}\BibitemShut {NoStop}%
\bibitem [{\citenamefont {Wolfgramm}\ \emph {et~al.}(2013)\citenamefont
  {Wolfgramm}, \citenamefont {Vitelli}, \citenamefont {Beduini}, \citenamefont
  {Godbout},\ and\ \citenamefont {Mitchell}}]{WolfgrammNP2013}%
  \BibitemOpen
  \bibfield  {author} {\bibinfo {author} {\bibfnamefont {F.}~\bibnamefont
  {Wolfgramm}}, \bibinfo {author} {\bibfnamefont {C.}~\bibnamefont {Vitelli}},
  \bibinfo {author} {\bibfnamefont {F.~A.}\ \bibnamefont {Beduini}}, \bibinfo
  {author} {\bibfnamefont {N.}~\bibnamefont {Godbout}},\ and\ \bibinfo {author}
  {\bibfnamefont {M.~W.}\ \bibnamefont {Mitchell}},\ }\bibfield  {title}
  {\bibinfo {title} {Entanglement-enhanced probing of a delicate material
  system},\ }\href {https://doi.org/10.1038/nphoton.2012.300} {\bibfield
  {journal} {\bibinfo  {journal} {Nat. Photonics}\ }\textbf {\bibinfo {volume}
  {7}},\ \bibinfo {pages} {28} (\bibinfo {year} {2013})}\BibitemShut {NoStop}%
\bibitem [{\citenamefont {Troullinou}(2022)}]{TroullinouThesis2022}%
  \BibitemOpen
  \bibfield  {author} {\bibinfo {author} {\bibfnamefont {C.}~\bibnamefont
  {Troullinou}},\ }\emph {\bibinfo {title} {Squeezed-ligh-enhanced magnetometry
  in a high density atomic vapor}},\ \href {http://hdl.handle.net/2117/379458}
  {Ph.D. thesis},\ \bibinfo  {school} {UPC, Institut de Ci{\`e}ncies
  Fot{\`o}niques} (\bibinfo {year} {2022})\BibitemShut {NoStop}%
\bibitem [{\citenamefont {Kitching}(2018)}]{KitchingAPR2018}%
  \BibitemOpen
  \bibfield  {author} {\bibinfo {author} {\bibfnamefont {J.}~\bibnamefont
  {Kitching}},\ }\bibfield  {title} {\bibinfo {title} {{Chip-scale atomic
  devices}},\ }\href {https://doi.org/10.1063/1.5026238} {\bibfield  {journal}
  {\bibinfo  {journal} {Appl. Phys. Rev.}\ }\textbf {\bibinfo {volume} {5}},\
  \bibinfo {pages} {031302} (\bibinfo {year} {2018})}\BibitemShut {NoStop}%
\bibitem [{\citenamefont {Deans}\ \emph {et~al.}(2018)\citenamefont {Deans},
  \citenamefont {Marmugi},\ and\ \citenamefont {Renzoni}}]{Deans2018}%
  \BibitemOpen
  \bibfield  {author} {\bibinfo {author} {\bibfnamefont {C.}~\bibnamefont
  {Deans}}, \bibinfo {author} {\bibfnamefont {L.}~\bibnamefont {Marmugi}},\
  and\ \bibinfo {author} {\bibfnamefont {F.}~\bibnamefont {Renzoni}},\
  }\bibfield  {title} {\bibinfo {title} {Active underwater detection with an
  array of atomic magnetometers},\ }\href@noop {} {\bibfield  {journal}
  {\bibinfo  {journal} {Applied optics}\ }\textbf {\bibinfo {volume} {57}},\
  \bibinfo {pages} {2346} (\bibinfo {year} {2018})}\BibitemShut {NoStop}%
\bibitem [{\citenamefont {Gerginov}\ \emph {et~al.}(2017)\citenamefont
  {Gerginov}, \citenamefont {Krzyzewski},\ and\ \citenamefont
  {Knappe}}]{Gerginov2017}%
  \BibitemOpen
  \bibfield  {author} {\bibinfo {author} {\bibfnamefont {V.}~\bibnamefont
  {Gerginov}}, \bibinfo {author} {\bibfnamefont {S.}~\bibnamefont
  {Krzyzewski}},\ and\ \bibinfo {author} {\bibfnamefont {S.}~\bibnamefont
  {Knappe}},\ }\bibfield  {title} {\bibinfo {title} {Pulsed operation of a
  miniature scalar optically pumped magnetometer},\ }\href
  {https://doi.org/10.1364/JOSAB.34.001429} {\bibfield  {journal} {\bibinfo
  {journal} {J. Opt. Soc. Am. B}\ }\textbf {\bibinfo {volume} {34}},\ \bibinfo
  {pages} {1429} (\bibinfo {year} {2017})}\BibitemShut {NoStop}%
\bibitem [{\citenamefont {{Defence Advanced Research Projects
  Agency}}(2016)}]{darpa_2016}%
  \BibitemOpen
  \bibfield  {author} {\bibinfo {author} {\bibnamefont {{Defence Advanced
  Research Projects Agency}}},\ }\href@noop {} {\bibinfo {title} {Underwater
  radio, anyone?}},\ \bibinfo {howpublished}
  {\url{https://www.darpa.mil/news-events/2016-12-16}} (\bibinfo {year}
  {2016}),\ \bibinfo {note} {accessed: 27/09/2022}\BibitemShut {NoStop}%
\bibitem [{\citenamefont {Page}\ \emph {et~al.}(2021)\citenamefont {Page},
  \citenamefont {Lambert}, \citenamefont {Mahmoudian}, \citenamefont {Newby},
  \citenamefont {Foley},\ and\ \citenamefont {Kornack}}]{page2021compact}%
  \BibitemOpen
  \bibfield  {author} {\bibinfo {author} {\bibfnamefont {B.~R.}\ \bibnamefont
  {Page}}, \bibinfo {author} {\bibfnamefont {R.}~\bibnamefont {Lambert}},
  \bibinfo {author} {\bibfnamefont {N.}~\bibnamefont {Mahmoudian}}, \bibinfo
  {author} {\bibfnamefont {D.~H.}\ \bibnamefont {Newby}}, \bibinfo {author}
  {\bibfnamefont {E.~L.}\ \bibnamefont {Foley}},\ and\ \bibinfo {author}
  {\bibfnamefont {T.~W.}\ \bibnamefont {Kornack}},\ }\bibfield  {title}
  {\bibinfo {title} {Compact quantum magnetometer system on an agile underwater
  glider},\ }\href@noop {} {\bibfield  {journal} {\bibinfo  {journal}
  {Sensors}\ }\textbf {\bibinfo {volume} {21}},\ \bibinfo {pages} {1092}
  (\bibinfo {year} {2021})}\BibitemShut {NoStop}%
\bibitem [{\citenamefont {Gibson}(2010)}]{gibson2010cave}%
  \BibitemOpen
  \bibfield  {author} {\bibinfo {author} {\bibfnamefont {D.}~\bibnamefont
  {Gibson}},\ }\href {https://books.google.pl/books?id=3o7FBgAAQBAJ} {\emph
  {\bibinfo {title} {Cave Radiolocation}}}\ (\bibinfo  {publisher} {Lulu.com},\
  \bibinfo {year} {2010})\BibitemShut {NoStop}%
\bibitem [{\citenamefont {Cohen}\ \emph {et~al.}(2009)\citenamefont {Cohen},
  \citenamefont {Inan},\ and\ \citenamefont {Paschal}}]{cohen_sensitive_2010}%
  \BibitemOpen
  \bibfield  {author} {\bibinfo {author} {\bibfnamefont {M.~B.}\ \bibnamefont
  {Cohen}}, \bibinfo {author} {\bibfnamefont {U.~S.}\ \bibnamefont {Inan}},\
  and\ \bibinfo {author} {\bibfnamefont {E.~W.}\ \bibnamefont {Paschal}},\
  }\bibfield  {title} {\bibinfo {title} {Sensitive broadband elf/vlf radio
  reception with the awesome instrument},\ }\href@noop {} {\bibfield  {journal}
  {\bibinfo  {journal} {IEEE Transactions on Geoscience and Remote Sensing}\
  }\textbf {\bibinfo {volume} {48}},\ \bibinfo {pages} {3} (\bibinfo {year}
  {2009})}\BibitemShut {NoStop}%
\bibitem [{\citenamefont {Fan}\ \emph {et~al.}(2022)\citenamefont {Fan},
  \citenamefont {Knappe},\ and\ \citenamefont {Gerginov}}]{Fan2022}%
  \BibitemOpen
  \bibfield  {author} {\bibinfo {author} {\bibfnamefont {I.}~\bibnamefont
  {Fan}}, \bibinfo {author} {\bibfnamefont {S.}~\bibnamefont {Knappe}},\ and\
  \bibinfo {author} {\bibfnamefont {V.}~\bibnamefont {Gerginov}},\ }\bibfield
  {title} {\bibinfo {title} {Magnetic communication by polarization helicity
  modulation using atomic magnetometers},\ }\href
  {https://doi.org/10.1063/5.0086169} {\bibfield  {journal} {\bibinfo
  {journal} {Rev. Sci. Instrum.}\ }\textbf {\bibinfo {volume} {93}},\ \bibinfo
  {pages} {053004} (\bibinfo {year} {2022})},\ \bibinfo {note} {publisher:
  American Institute of Physics}\BibitemShut {NoStop}%
\bibitem [{\citenamefont {Savukov}\ \emph {et~al.}(2007)\citenamefont
  {Savukov}, \citenamefont {Seltzer},\ and\ \citenamefont
  {Romalis}}]{SavukovJMR2007}%
  \BibitemOpen
  \bibfield  {author} {\bibinfo {author} {\bibfnamefont {I.}~\bibnamefont
  {Savukov}}, \bibinfo {author} {\bibfnamefont {S.}~\bibnamefont {Seltzer}},\
  and\ \bibinfo {author} {\bibfnamefont {M.}~\bibnamefont {Romalis}},\
  }\bibfield  {title} {\bibinfo {title} {Detection of nmr signals with a
  radio-frequency atomic magnetometer},\ }\href
  {https://doi.org/https://doi.org/10.1016/j.jmr.2006.12.012} {\bibfield
  {journal} {\bibinfo  {journal} {J. Magn. Reson.}\ }\textbf {\bibinfo {volume}
  {185}},\ \bibinfo {pages} {214} (\bibinfo {year} {2007})}\BibitemShut
  {NoStop}%
\bibitem [{\citenamefont {Ingleby}\ \emph {et~al.}(2020)\citenamefont
  {Ingleby}, \citenamefont {Chalmers}, \citenamefont {Dyer}, \citenamefont
  {Griffin},\ and\ \citenamefont {Riis}}]{Ingleby2020}%
  \BibitemOpen
  \bibfield  {author} {\bibinfo {author} {\bibfnamefont {S.~J.}\ \bibnamefont
  {Ingleby}}, \bibinfo {author} {\bibfnamefont {I.~C.}\ \bibnamefont
  {Chalmers}}, \bibinfo {author} {\bibfnamefont {T.~E.}\ \bibnamefont {Dyer}},
  \bibinfo {author} {\bibfnamefont {P.~F.}\ \bibnamefont {Griffin}},\ and\
  \bibinfo {author} {\bibfnamefont {E.}~\bibnamefont {Riis}},\ }\href
  {https://doi.org/10.48550/arXiv.2003.03267} {\bibinfo {title} {Resonant very
  low- and ultra low frequency digital signal reception using a portable atomic
  magnetometer}} (\bibinfo {year} {2020}),\ \Eprint
  {https://arxiv.org/abs/2003.03267} {arXiv:2003.03267 [physics.atom-ph]}
  \BibitemShut {NoStop}%
\bibitem [{\citenamefont {Funaki}\ \emph {et~al.}(2014)\citenamefont {Funaki},
  \citenamefont {Higashino}, \citenamefont {Sakanaka}, \citenamefont {Iwata},
  \citenamefont {Nakamura}, \citenamefont {Hirasawa}, \citenamefont {Obara},\
  and\ \citenamefont {Kuwabara}}]{FunakiPS2014}%
  \BibitemOpen
  \bibfield  {author} {\bibinfo {author} {\bibfnamefont {M.}~\bibnamefont
  {Funaki}}, \bibinfo {author} {\bibfnamefont {S.-I.}\ \bibnamefont
  {Higashino}}, \bibinfo {author} {\bibfnamefont {S.}~\bibnamefont {Sakanaka}},
  \bibinfo {author} {\bibfnamefont {N.}~\bibnamefont {Iwata}}, \bibinfo
  {author} {\bibfnamefont {N.}~\bibnamefont {Nakamura}}, \bibinfo {author}
  {\bibfnamefont {N.}~\bibnamefont {Hirasawa}}, \bibinfo {author}
  {\bibfnamefont {N.}~\bibnamefont {Obara}},\ and\ \bibinfo {author}
  {\bibfnamefont {M.}~\bibnamefont {Kuwabara}},\ }\bibfield  {title} {\bibinfo
  {title} {Small unmanned aerial vehicles for aeromagnetic surveys and their
  flights in the {South} {Shetland} {Islands}, {Antarctica}},\ }\href
  {https://doi.org/10.1016/j.polar.2014.07.001} {\bibfield  {journal} {\bibinfo
   {journal} {Polar Sci.}\ }\textbf {\bibinfo {volume} {8}},\ \bibinfo {pages}
  {342} (\bibinfo {year} {2014})}\BibitemShut {NoStop}%
\bibitem [{\citenamefont {Becken}\ \emph {et~al.}(2022)\citenamefont {Becken},
  \citenamefont {Kotowski}, \citenamefont {Schmalzl}, \citenamefont {Symons},\
  and\ \citenamefont {Brauch}}]{becken2022semi}%
  \BibitemOpen
  \bibfield  {author} {\bibinfo {author} {\bibfnamefont {M.}~\bibnamefont
  {Becken}}, \bibinfo {author} {\bibfnamefont {P.~O.}\ \bibnamefont
  {Kotowski}}, \bibinfo {author} {\bibfnamefont {J.}~\bibnamefont {Schmalzl}},
  \bibinfo {author} {\bibfnamefont {G.}~\bibnamefont {Symons}},\ and\ \bibinfo
  {author} {\bibfnamefont {K.}~\bibnamefont {Brauch}},\ }\bibfield  {title}
  {\bibinfo {title} {Semi-airborne electromagnetic exploration using a scalar
  magnetometer suspended below a multicopter},\ }\href@noop {} {\bibfield
  {journal} {\bibinfo  {journal} {First Break}\ }\textbf {\bibinfo {volume}
  {40}},\ \bibinfo {pages} {37} (\bibinfo {year} {2022})}\BibitemShut {NoStop}%
\bibitem [{\citenamefont {Kemp}\ \emph {et~al.}(2019)\citenamefont {Kemp},
  \citenamefont {Franzi}, \citenamefont {Haase}, \citenamefont {Jongewaard},
  \citenamefont {Whittaker}, \citenamefont {Kirkpatrick},\ and\ \citenamefont
  {Sparr}}]{Kemp2019}%
  \BibitemOpen
  \bibfield  {author} {\bibinfo {author} {\bibfnamefont {M.~A.}\ \bibnamefont
  {Kemp}}, \bibinfo {author} {\bibfnamefont {M.}~\bibnamefont {Franzi}},
  \bibinfo {author} {\bibfnamefont {A.}~\bibnamefont {Haase}}, \bibinfo
  {author} {\bibfnamefont {E.}~\bibnamefont {Jongewaard}}, \bibinfo {author}
  {\bibfnamefont {M.~T.}\ \bibnamefont {Whittaker}}, \bibinfo {author}
  {\bibfnamefont {M.}~\bibnamefont {Kirkpatrick}},\ and\ \bibinfo {author}
  {\bibfnamefont {R.}~\bibnamefont {Sparr}},\ }\bibfield  {title} {\bibinfo
  {title} {A high q piezoelectric resonator as a portable vlf transmitter},\
  }\href {https://doi.org/10.1038/s41467-019-09680-2} {\bibfield  {journal}
  {\bibinfo  {journal} {Nat. Commun.}\ }\textbf {\bibinfo {volume} {10}},\
  \bibinfo {pages} {1715} (\bibinfo {year} {2019})}\BibitemShut {NoStop}%
\bibitem [{\citenamefont {Kong}\ \emph {et~al.}(2020)\citenamefont {Kong},
  \citenamefont {Jim{\'e}nez-Mart{\'\i}nez}, \citenamefont {Troullinou},
  \citenamefont {Lucivero}, \citenamefont {T{\'o}th},\ and\ \citenamefont
  {Mitchell}}]{KongNC2020}%
  \BibitemOpen
  \bibfield  {author} {\bibinfo {author} {\bibfnamefont {J.}~\bibnamefont
  {Kong}}, \bibinfo {author} {\bibfnamefont {R.}~\bibnamefont
  {Jim{\'e}nez-Mart{\'\i}nez}}, \bibinfo {author} {\bibfnamefont
  {C.}~\bibnamefont {Troullinou}}, \bibinfo {author} {\bibfnamefont {V.~G.}\
  \bibnamefont {Lucivero}}, \bibinfo {author} {\bibfnamefont {G.}~\bibnamefont
  {T{\'o}th}},\ and\ \bibinfo {author} {\bibfnamefont {M.~W.}\ \bibnamefont
  {Mitchell}},\ }\bibfield  {title} {\bibinfo {title} {Measurement-induced,
  spatially-extended entanglement in a hot, strongly-interacting atomic
  system},\ }\href {https://doi.org/10.1038/s41467-020-15899-1} {\bibfield
  {journal} {\bibinfo  {journal} {Nat. Commun.}\ }\textbf {\bibinfo {volume}
  {11}},\ \bibinfo {pages} {2415} (\bibinfo {year} {2020})}\BibitemShut
  {NoStop}%
\bibitem [{\citenamefont {Caves}(1981{\natexlab{b}})}]{PhysRevD.23.1693}%
  \BibitemOpen
  \bibfield  {author} {\bibinfo {author} {\bibfnamefont {C.~M.}\ \bibnamefont
  {Caves}},\ }\bibfield  {title} {\bibinfo {title} {Quantum-mechanical noise in
  an interferometer},\ }\href {https://doi.org/10.1103/PhysRevD.23.1693}
  {\bibfield  {journal} {\bibinfo  {journal} {Phys. Rev. D}\ }\textbf {\bibinfo
  {volume} {23}},\ \bibinfo {pages} {1693} (\bibinfo {year}
  {1981}{\natexlab{b}})}\BibitemShut {NoStop}%
\bibitem [{\citenamefont {Colangelo}\ \emph
  {et~al.}(2017{\natexlab{c}})\citenamefont {Colangelo}, \citenamefont
  {Martin~Ciurana}, \citenamefont {Puentes}, \citenamefont {Mitchell},\ and\
  \citenamefont {Sewell}}]{ColangeloPRL2017}%
  \BibitemOpen
  \bibfield  {author} {\bibinfo {author} {\bibfnamefont {G.}~\bibnamefont
  {Colangelo}}, \bibinfo {author} {\bibfnamefont {F.}~\bibnamefont
  {Martin~Ciurana}}, \bibinfo {author} {\bibfnamefont {G.}~\bibnamefont
  {Puentes}}, \bibinfo {author} {\bibfnamefont {M.~W.}\ \bibnamefont
  {Mitchell}},\ and\ \bibinfo {author} {\bibfnamefont {R.~J.}\ \bibnamefont
  {Sewell}},\ }\bibfield  {title} {\bibinfo {title} {Entanglement-enhanced
  phase estimation without prior phase information},\ }\href
  {https://doi.org/10.1103/PhysRevLett.118.233603} {\bibfield  {journal}
  {\bibinfo  {journal} {Phys. Rev. Lett.}\ }\textbf {\bibinfo {volume} {118}},\
  \bibinfo {pages} {233603} (\bibinfo {year} {2017}{\natexlab{c}})}\BibitemShut
  {NoStop}%
\bibitem [{\citenamefont {Shah}\ \emph {et~al.}(2010)\citenamefont {Shah},
  \citenamefont {Vasilakis},\ and\ \citenamefont {Romalis}}]{Shah2010}%
  \BibitemOpen
  \bibfield  {author} {\bibinfo {author} {\bibfnamefont {V.}~\bibnamefont
  {Shah}}, \bibinfo {author} {\bibfnamefont {G.}~\bibnamefont {Vasilakis}},\
  and\ \bibinfo {author} {\bibfnamefont {M.~V.}\ \bibnamefont {Romalis}},\
  }\bibfield  {title} {\bibinfo {title} {High bandwidth atomic magnetometery
  with continuous quantum nondemolition measurements},\ }\href
  {https://doi.org/10.1103/PhysRevLett.104.013601} {\bibfield  {journal}
  {\bibinfo  {journal} {Phys. Rev. Lett.}\ }\textbf {\bibinfo {volume} {104}},\
  \bibinfo {pages} {013601} (\bibinfo {year} {2010})}\BibitemShut {NoStop}%
\bibitem [{\citenamefont {Vasilakis}\ \emph {et~al.}(2015)\citenamefont
  {Vasilakis}, \citenamefont {Shen}, \citenamefont {Jensen}, \citenamefont
  {Balabas}, \citenamefont {Salart}, \citenamefont {Chen},\ and\ \citenamefont
  {Polzik}}]{Vasilakis2015}%
  \BibitemOpen
  \bibfield  {author} {\bibinfo {author} {\bibfnamefont {G.}~\bibnamefont
  {Vasilakis}}, \bibinfo {author} {\bibfnamefont {H.}~\bibnamefont {Shen}},
  \bibinfo {author} {\bibfnamefont {K.}~\bibnamefont {Jensen}}, \bibinfo
  {author} {\bibfnamefont {M.}~\bibnamefont {Balabas}}, \bibinfo {author}
  {\bibfnamefont {D.}~\bibnamefont {Salart}}, \bibinfo {author} {\bibfnamefont
  {B.}~\bibnamefont {Chen}},\ and\ \bibinfo {author} {\bibfnamefont {E.~S.}\
  \bibnamefont {Polzik}},\ }\bibfield  {title} {\bibinfo {title} {Generation of
  a squeezed state of an oscillator by stroboscopic back-action-evading
  measurement},\ }\href {http://dx.doi.org/10.1038/nphys3280} {\bibfield
  {journal} {\bibinfo  {journal} {Nat. Phys}\ }\textbf {\bibinfo {volume}
  {11}},\ \bibinfo {pages} {389} (\bibinfo {year} {2015})}\BibitemShut
  {NoStop}%
\bibitem [{\citenamefont {Bao}\ \emph {et~al.}(2020)\citenamefont {Bao},
  \citenamefont {Duan}, \citenamefont {Jin}, \citenamefont {Lu}, \citenamefont
  {Li}, \citenamefont {Qu}, \citenamefont {Wang}, \citenamefont {Novikova},
  \citenamefont {Mikhailov}, \citenamefont {Zhao}, \citenamefont {Mølmer},
  \citenamefont {Shen},\ and\ \citenamefont {Xiao}}]{bao2020}%
  \BibitemOpen
  \bibfield  {author} {\bibinfo {author} {\bibfnamefont {H.}~\bibnamefont
  {Bao}}, \bibinfo {author} {\bibfnamefont {J.}~\bibnamefont {Duan}}, \bibinfo
  {author} {\bibfnamefont {S.}~\bibnamefont {Jin}}, \bibinfo {author}
  {\bibfnamefont {X.}~\bibnamefont {Lu}}, \bibinfo {author} {\bibfnamefont
  {P.}~\bibnamefont {Li}}, \bibinfo {author} {\bibfnamefont {W.}~\bibnamefont
  {Qu}}, \bibinfo {author} {\bibfnamefont {M.}~\bibnamefont {Wang}}, \bibinfo
  {author} {\bibfnamefont {I.}~\bibnamefont {Novikova}}, \bibinfo {author}
  {\bibfnamefont {E.~E.}\ \bibnamefont {Mikhailov}}, \bibinfo {author}
  {\bibfnamefont {K.-F.}\ \bibnamefont {Zhao}}, \bibinfo {author}
  {\bibfnamefont {K.}~\bibnamefont {Mølmer}}, \bibinfo {author} {\bibfnamefont
  {H.}~\bibnamefont {Shen}},\ and\ \bibinfo {author} {\bibfnamefont
  {Y.}~\bibnamefont {Xiao}},\ }\bibfield  {title} {\bibinfo {title} {Spin
  squeezing of $10^{11}$ atoms by prediction and retrodiction measurements},\
  }\href {https://doi.org/10.1038/s41586-020-2243-7} {\bibfield  {journal}
  {\bibinfo  {journal} {Nature}\ }\textbf {\bibinfo {volume} {581}},\ \bibinfo
  {pages} {159} (\bibinfo {year} {2020})}\BibitemShut {NoStop}%
\bibitem [{\citenamefont {Guarrera}\ \emph {et~al.}(2019)\citenamefont
  {Guarrera}, \citenamefont {Gartman}, \citenamefont {Bevilacqua},
  \citenamefont {Barontini},\ and\ \citenamefont
  {Chalupczak}}]{guarrera2019parametric}%
  \BibitemOpen
  \bibfield  {author} {\bibinfo {author} {\bibfnamefont {V.}~\bibnamefont
  {Guarrera}}, \bibinfo {author} {\bibfnamefont {R.}~\bibnamefont {Gartman}},
  \bibinfo {author} {\bibfnamefont {G.}~\bibnamefont {Bevilacqua}}, \bibinfo
  {author} {\bibfnamefont {G.}~\bibnamefont {Barontini}},\ and\ \bibinfo
  {author} {\bibfnamefont {W.}~\bibnamefont {Chalupczak}},\ }\bibfield  {title}
  {\bibinfo {title} {Parametric amplification and noise squeezing in room
  temperature atomic vapors},\ }\href@noop {} {\bibfield  {journal} {\bibinfo
  {journal} {Phys. Rev. Lett.}\ }\textbf {\bibinfo {volume} {123}},\ \bibinfo
  {pages} {033601} (\bibinfo {year} {2019})}\BibitemShut {NoStop}%
\bibitem [{\citenamefont {Guarrera}\ \emph {et~al.}(2021)\citenamefont
  {Guarrera}, \citenamefont {Gartman}, \citenamefont {Bevilacqua},\ and\
  \citenamefont {Chalupczak}}]{guarrera2021spin}%
  \BibitemOpen
  \bibfield  {author} {\bibinfo {author} {\bibfnamefont {V.}~\bibnamefont
  {Guarrera}}, \bibinfo {author} {\bibfnamefont {R.}~\bibnamefont {Gartman}},
  \bibinfo {author} {\bibfnamefont {G.}~\bibnamefont {Bevilacqua}},\ and\
  \bibinfo {author} {\bibfnamefont {W.}~\bibnamefont {Chalupczak}},\ }\bibfield
   {title} {\bibinfo {title} {Spin-noise spectroscopy of a noise-squeezed
  atomic state},\ }\href@noop {} {\bibfield  {journal} {\bibinfo  {journal}
  {Phys. rev. res.}\ }\textbf {\bibinfo {volume} {3}},\ \bibinfo {pages}
  {L032015} (\bibinfo {year} {2021})}\BibitemShut {NoStop}%
\bibitem [{\citenamefont {Martin~Ciurana}\ \emph {et~al.}(2017)\citenamefont
  {Martin~Ciurana}, \citenamefont {Colangelo}, \citenamefont {Slodička},
  \citenamefont {Sewell},\ and\ \citenamefont {Mitchell}}]{MartinCiurana2017}%
  \BibitemOpen
  \bibfield  {author} {\bibinfo {author} {\bibfnamefont {F.}~\bibnamefont
  {Martin~Ciurana}}, \bibinfo {author} {\bibfnamefont {G.}~\bibnamefont
  {Colangelo}}, \bibinfo {author} {\bibfnamefont {L.}~\bibnamefont
  {Slodička}}, \bibinfo {author} {\bibfnamefont {R.~J.}\ \bibnamefont
  {Sewell}},\ and\ \bibinfo {author} {\bibfnamefont {M.~W.}\ \bibnamefont
  {Mitchell}},\ }\bibfield  {title} {\bibinfo {title} {Entanglement-enhanced
  radio-frequency field detection and waveform sensing},\ }\href
  {https://doi.org/10.1103/PhysRevLett.119.043603} {\bibfield  {journal}
  {\bibinfo  {journal} {Phys. Rev. Lett.}\ }\textbf {\bibinfo {volume} {119}},\
  \bibinfo {pages} {043603} (\bibinfo {year} {2017})}\BibitemShut {NoStop}%
\bibitem [{\citenamefont {Bell}\ and\ \citenamefont
  {Bloom}(1961)}]{bell1961optically}%
  \BibitemOpen
  \bibfield  {author} {\bibinfo {author} {\bibfnamefont {W.~E.}\ \bibnamefont
  {Bell}}\ and\ \bibinfo {author} {\bibfnamefont {A.~L.}\ \bibnamefont
  {Bloom}},\ }\bibfield  {title} {\bibinfo {title} {Optically driven spin
  precession},\ }\href@noop {} {\bibfield  {journal} {\bibinfo  {journal}
  {Phys. Rev. Lett.}\ }\textbf {\bibinfo {volume} {6}},\ \bibinfo {pages} {280}
  (\bibinfo {year} {1961})}\BibitemShut {NoStop}%
\bibitem [{\citenamefont {Mouloudakis}\ \emph {et~al.}(2022)\citenamefont
  {Mouloudakis}, \citenamefont {Vasilakis}, \citenamefont {Lucivero},
  \citenamefont {Kong}, \citenamefont {Kominis},\ and\ \citenamefont
  {Mitchell}}]{MouloudakisPRA2022}%
  \BibitemOpen
  \bibfield  {author} {\bibinfo {author} {\bibfnamefont {K.}~\bibnamefont
  {Mouloudakis}}, \bibinfo {author} {\bibfnamefont {G.}~\bibnamefont
  {Vasilakis}}, \bibinfo {author} {\bibfnamefont {V.~G.}\ \bibnamefont
  {Lucivero}}, \bibinfo {author} {\bibfnamefont {J.}~\bibnamefont {Kong}},
  \bibinfo {author} {\bibfnamefont {I.~K.}\ \bibnamefont {Kominis}},\ and\
  \bibinfo {author} {\bibfnamefont {M.~W.}\ \bibnamefont {Mitchell}},\
  }\bibfield  {title} {\bibinfo {title} {Effects of spin-exchange collisions on
  the fluctuation spectra of hot alkali-metal vapors},\ }\href
  {https://doi.org/10.1103/PhysRevA.106.023112} {\bibfield  {journal} {\bibinfo
   {journal} {Phys. Rev. A}\ }\textbf {\bibinfo {volume} {106}},\ \bibinfo
  {pages} {023112} (\bibinfo {year} {2022})}\BibitemShut {NoStop}%
\bibitem [{\citenamefont {Troullinou}\ \emph
  {et~al.}(2021{\natexlab{b}})\citenamefont {Troullinou}, \citenamefont
  {Jim{\'e}nez-Mart{\'\i}nez}, \citenamefont {Kong}, \citenamefont {Lucivero},\
  and\ \citenamefont {Mitchell}}]{troullinou2021squeezed}%
  \BibitemOpen
  \bibfield  {author} {\bibinfo {author} {\bibfnamefont {C.}~\bibnamefont
  {Troullinou}}, \bibinfo {author} {\bibfnamefont {R.}~\bibnamefont
  {Jim{\'e}nez-Mart{\'\i}nez}}, \bibinfo {author} {\bibfnamefont
  {J.}~\bibnamefont {Kong}}, \bibinfo {author} {\bibfnamefont {V.}~\bibnamefont
  {Lucivero}},\ and\ \bibinfo {author} {\bibfnamefont {M.}~\bibnamefont
  {Mitchell}},\ }\bibfield  {title} {\bibinfo {title} {Squeezed-light
  enhancement and backaction evasion in a high sensitivity optically pumped
  magnetometer},\ }\href@noop {} {\bibfield  {journal} {\bibinfo  {journal}
  {Phys. Rev. Lett.}\ }\textbf {\bibinfo {volume} {127}},\ \bibinfo {pages}
  {193601} (\bibinfo {year} {2021}{\natexlab{b}})}\BibitemShut {NoStop}%
\bibitem [{\citenamefont {{Gerginov}}\ \emph {et~al.}(2020)\citenamefont
  {{Gerginov}}, \citenamefont {{Pomponio}},\ and\ \citenamefont
  {{Knappe}}}]{Gerginov2020}%
  \BibitemOpen
  \bibfield  {author} {\bibinfo {author} {\bibfnamefont {V.}~\bibnamefont
  {{Gerginov}}}, \bibinfo {author} {\bibfnamefont {M.}~\bibnamefont
  {{Pomponio}}},\ and\ \bibinfo {author} {\bibfnamefont {S.}~\bibnamefont
  {{Knappe}}},\ }\bibfield  {title} {\bibinfo {title} {Scalar magnetometry
  below 100 ft/hz1/2 in a microfabricated cell},\ }\href
  {https://doi.org/10.1109/JSEN.2020.3002193} {\bibfield  {journal} {\bibinfo
  {journal} {IEEE Sens. J.}\ }\textbf {\bibinfo {volume} {20}},\ \bibinfo
  {pages} {12684} (\bibinfo {year} {2020})}\BibitemShut {NoStop}%
\bibitem [{\citenamefont {Jimenez-Martinez}\ \emph {et~al.}(2010)\citenamefont
  {Jimenez-Martinez}, \citenamefont {Griffith}, \citenamefont {Wang},
  \citenamefont {Knappe}, \citenamefont {Kitching}, \citenamefont {Smith},\
  and\ \citenamefont {Prouty}}]{Jimenez-Martinez2010}%
  \BibitemOpen
  \bibfield  {author} {\bibinfo {author} {\bibfnamefont {R.}~\bibnamefont
  {Jimenez-Martinez}}, \bibinfo {author} {\bibfnamefont {W.~C.}\ \bibnamefont
  {Griffith}}, \bibinfo {author} {\bibfnamefont {Y.}~\bibnamefont {Wang}},
  \bibinfo {author} {\bibfnamefont {S.}~\bibnamefont {Knappe}}, \bibinfo
  {author} {\bibfnamefont {J.}~\bibnamefont {Kitching}}, \bibinfo {author}
  {\bibfnamefont {K.}~\bibnamefont {Smith}},\ and\ \bibinfo {author}
  {\bibfnamefont {M.~D.}\ \bibnamefont {Prouty}},\ }\bibfield  {title}
  {\bibinfo {title} {Sensitivity comparison of {Mx} and frequency-modulated
  bell--bloom {Cs} magnetometers in a microfabricated cell},\ }\href
  {https://doi.org/10.1109/TIM.2009.2023829} {\bibfield  {journal} {\bibinfo
  {journal} {IEEE Trans. Instrum. Meas.}\ }\textbf {\bibinfo {volume} {59}},\
  \bibinfo {pages} {372} (\bibinfo {year} {2010})}\BibitemShut {NoStop}%
\bibitem [{Rep(2023)}]{RepositoryData}%
  \BibitemOpen
  \href {https://doi.org/https://doi.org/10.7910/DVN/9MEVFW} {\bibinfo {title}
  {Data for: Multi-parameter quantum sensing and magnetic communications with a
  hybrid dc/rf optically-pumped magnetometer}} (\bibinfo {year}
  {2023})\BibitemShut {NoStop}%
\end{thebibliography}%

\end{document}